\documentclass[10pt,twocolumn,letterpaper]{article}

\usepackage[pagenumbers]{cvpr} 

\definecolor{cvprblue}{rgb}{0.21,0.49,0.74}
\usepackage[pagebackref,breaklinks,colorlinks,allcolors=cvprblue]{hyperref}

\usepackage{natbib}
\usepackage{float}
\usepackage{multirow}
\usepackage{makecell}
\usepackage{graphicx}

\usepackage{algorithm}
\usepackage{listings}
\usepackage{amsmath}
\usepackage{amsfonts}
\usepackage{tikz}

\usepackage{pifont}
\usepackage{color}
\usepackage{colortbl}

\usepackage{multicol}
\usepackage{xspace}
\newcommand{\cmark}{\ding{52}\xspace}%
\newcommand{\xmarkg}{\textcolor{lightgray}{\ding{56}}\xspace}%
\definecolor{bluee}{RGB}{225, 235, 246}
\definecolor{upup}{RGB}{255,0,0}
\definecolor{down}{RGB}{83,100,147}

\def\paperID{14233} 
\def\confName{CVPR}
\def\confYear{2025}

\title{A Synergy Scoring Filter for Unsupervised Anomaly Detection with Noisy Data}

\author{
Chengming Liu \quad
Fengjie Wang \quad
Lei Shi \quad
Zhe Zhao \\
Zhengzhou University \\
\texttt{cmliu@zzu.edu.cn},\quad\texttt{oo.ggg42@gmail.com} \\
\texttt{shilei@zzu.edu.cn},\quad\texttt{sevenzz@zzu.edu.cn}}

\begin{document}
\maketitle
\begin{abstract}
    Noise-inclusive fully unsupervised anomaly detection (FUAD) holds significant practical relevance. Although various methods exist to address this problem, they are limited in both performance and scalability. Our work seeks to overcome these obstacles, enabling broader adaptability of unsupervised anomaly detection (UAD) models to FUAD. To achieve this, we introduce the Synergy Scoring Filter (SSFilter), the first fully unsupervised anomaly detection approach to leverage sample-level filtering. SSFilter facilitates end-to-end robust training and applies filtering to the complete training set post-training, offering a model-agnostic solution for FUAD. Specifically, SSFilter integrates a batch-level anomaly scoring mechanism based on mutual patch comparison and utilizes regression errors in anomalous regions, alongside prediction uncertainty, to estimate sample-level uncertainty scores that calibrate the anomaly scoring mechanism. This design produces a synergistic, robust filtering approach. Furthermore, we propose a realistic anomaly synthesis method and an integrity enhancement strategy to improve model training and mitigate missed noisy samples. Our method establishes state-of-the-art performance on the FUAD benchmark of the recent large-scale industrial anomaly detection dataset, Real-IAD. Additionally, dataset-level filtering enhances the performance of various UAD methods on the FUAD benchmark, and the high scalability of our approach significantly boosts its practical applicability.
\end{abstract}

\section{Introduction}
\label{sec:intro}

\begin{figure}[t]
  \centering
  \includegraphics[width=1\linewidth]{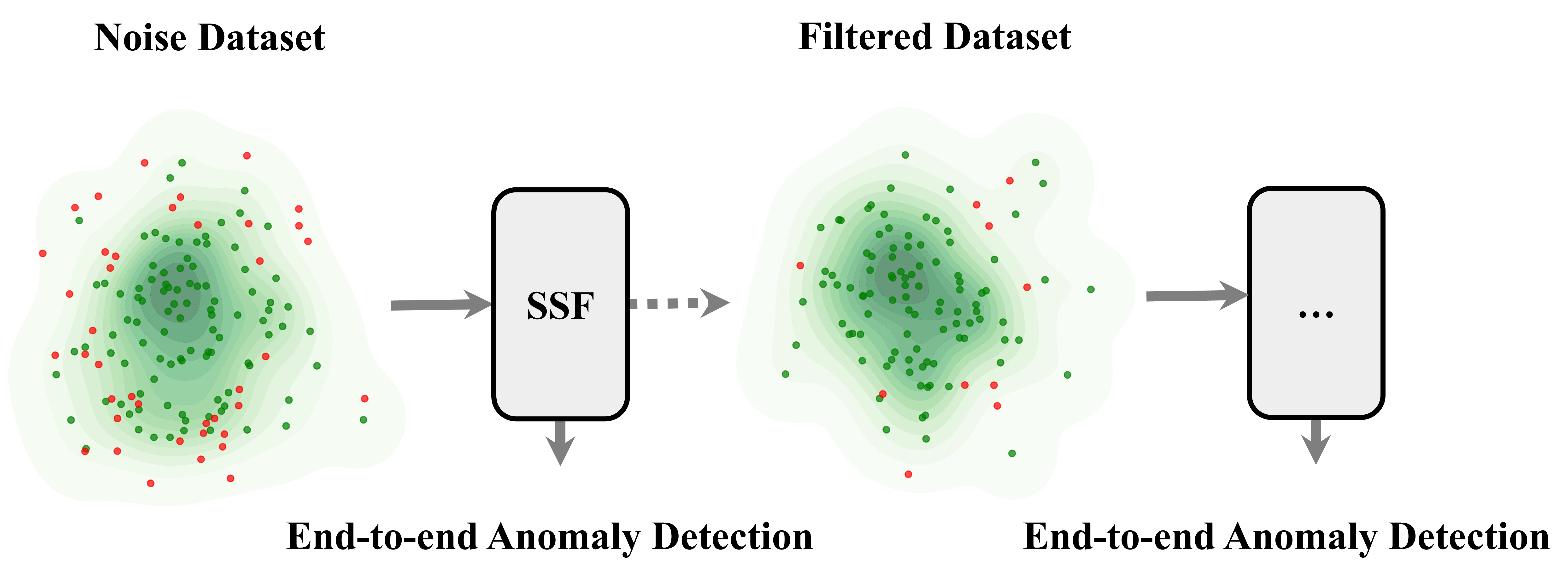}
   \caption{ Filter the entire dataset using the trained SSFilter, extending to various other UAD methods.}
   \label{fig:dataset}
\end{figure}

Image anomaly detection can be broadly categorized into semantic anomaly detection and sensory anomaly detection \cite{yangGeneralizedOutDistributionDetection2024}. Semantic anomaly detection emphasizes the overall semantics of the image, while sensory anomaly detection focuses on fine-grained details, such as small scratches on objects. This paper concentrates on sensory anomaly detection, which is already widely used in fields like intelligent manufacturing, lesion detection, and video surveillance. Due to the difficulty of obtaining anomalous samples, the unsupervised paradigm of training with only normal samples has gained great favor. However, as the scale increases, selecting a perfectly normal sample dataset still requires significant labor and time costs. Additionally, missed anomalies during screening can degrade model performance. Therefore, accounting for a certain level of noisy samples within the training set holds substantial practical relevance. Moreover, in sensory anomaly detection, the high similarity between normal and abnormal samples makes the subtle sensory contamination fundamentally different from the more pronounced semantic contamination commonly addressed in conventional noisy learning.

Existing unsupervised anomaly detection methods are generally categorized into three mainstream approaches: reconstruction-based methods \cite{dengAnomalyDetectionReverse2022,zhangIndustrialAnomalyDetection2023, guoDinomalyLessMore2024, wangMiniMaxADLightweightAutoencoder2024, heMambaADExploringState2024}, anomaly synthesizing-based methods \cite{zavrtanikDRAEMDiscriminativelyTrained2021,liuSimpleNetSimpleNetwork2023,zhangDeSTSegSegmentationGuided2023, yangSLSGIndustrialImage2023}, and embedding-based methods \cite{rothTotalRecallIndustrial2022, defardPaDiMPatchDistribution2021}. Since these methods assume a clean training set, their performance degrades when confronted with noisy data. PatchCore \cite{rothTotalRecallIndustrial2022} is a foundational approach that extracts patch-level feature embeddings of normal images into a Memory Bank, detecting anomalous patches during inference through a patch query process. Recently, SoftPatch \cite{jiangSoftPatchUnsupervisedAnomaly2022}, building on PatchCore \cite{rothTotalRecallIndustrial2022}, introduced a patch-level filtering strategy where patch features are filtered and weighted before being stored in the Memory Bank to reduce contamination from anomalous patches, thereby enhancing model robustness. However, the improvement in performance remains limited, and the proposed patch-level filtering strategy is not easily extendable to other methods.

To address this issue, we developed a highly scalable and high-performance robust training method with strong sample-level filtering capabilities. This method can independently conduct end-to-end training and directly perform sample-level filtering on the training set post-training, allowing other methods to benefit easily, as illustrated in the overall concept diagram in Figure \ref{fig:dataset}. Specifically, drawing inspiration from advanced zero-shot batch-level anomaly detection techniques \cite{liMuScZeroshotIndustrial2024, dammAnomalyDINOBoostingPatchbased2024a}, we designed an efficient in-batch sample filtering mechanism. The core principle is that anomalous patches exhibit higher sparsity and specificity, whereas normal patches are densely distributed and interconnected, enabling mutual scoring of patches. Anomalous patches, which lack similar counterparts, receive higher anomaly scores. This approach allows us to obtain anomaly scores for samples within the batch, filtering out high-risk samples to mitigate the noise's impact on the model during iterative training.

However, this scoring strategy, as a zero-shot approach without training, introduces inherent biases. To mitigate the model's over-reliance on such biases, we incorporate uncertainty estimation \cite{galDropoutBayesianApproximation2016} and perform fine-grained, sample-level uncertainty estimation based on regression (reconstruction) errors and prediction uncertainty in anomalous regions. This uncertainty score serves as a corrective factor in the filtering process. Additionally, to address inevitable omissions, we developed a realistic anomaly synthesis method using by-products from uncertainty estimation and implemented a ``restoring the hidden truth'' strategy to guide model training and counteract anomalous interference.

Our method achieved state-of-the-art (SOTA) performance on the fully unsupervised benchmark of the latest large-scale industrial anomaly detection dataset, Real-IAD \cite{wangRealIADRealworldMultiview2024}. Additionally, our dataset-level filtering approach enhances fully unsupervised performance across various methods, supporting effective method selection for real-world applications. In summary, our contributions are as follows:

\begin{itemize}
  \item To the best of our knowledge, our proposed SSFilter method is the first sample-level filtering-based approach for fully unsupervised anomaly detection. Its exceptional scalability and performance significantly bridge the gap between fully unsupervised anomaly detection and traditional unsupervised anomaly detection.
  \item We proposed a sample-level filtering strategy based on the synergy of patch mutual scoring and uncertainty estimation, which performs excellently.
  \item We proposed a realistic anomaly synthesis method, upon which we built a ``restoring the hidden truth'' strategy to enhance model performance.
\end{itemize}

\section{Related Work}
\label{sec:related}

\begin{figure*}[t]
  \centering
  \includegraphics[width=1.0\linewidth]{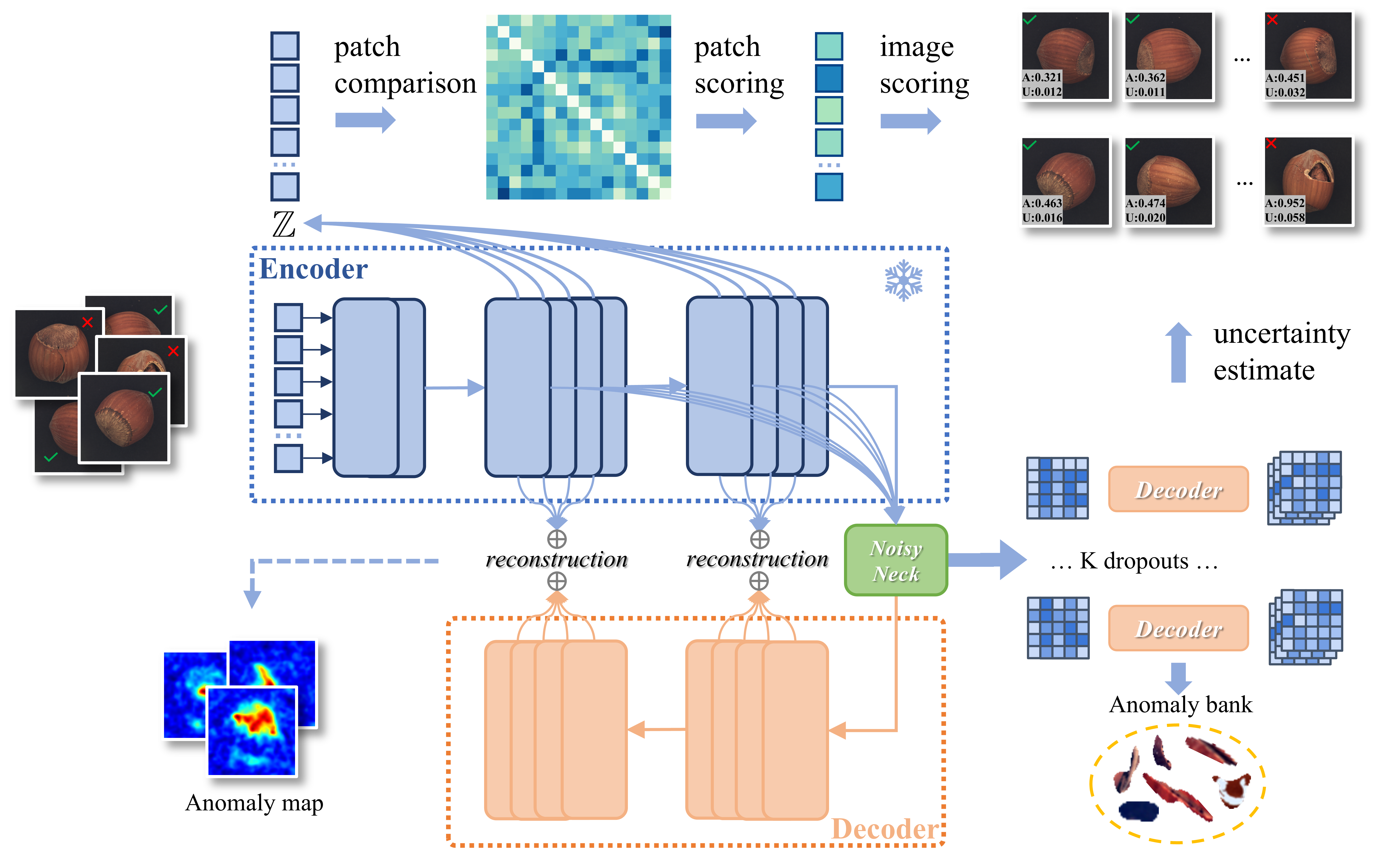}
   \caption{The overall framework of SSFilter. The patch comparison scorer and uncertainty estimator work synergistically to filter high-risk samples in the batch.} 
   \label{fig:framework}
\end{figure*}
\subsection{Unsupervised Anomaly Detection}
\textbf{Reconstruction.}
Reconstruction-based methods operate on a fundamental assumption: during training, normal images are accurately reconstructed, while during inference, anomalous regions fail to reconstruct due to the model's lack of exposure to these patterns. Early models focused on pixel-level reconstruction, but they often produced suboptimal results \cite{bergmannImprovingUnsupervisedDefect2019a,gongMemorizingNormalityDetect2019a,liuVisuallyExplainingVariational2020,parkLearningMemoryguidedNormality2020}. Recently, researchers have aimed to enhance reconstruction quality by decoupling image frequencies \cite{liangOmniFrequencyChannelSelectionRepresentations2023} or employing advanced diffusion models \cite{mousakhanAnomalyDetectionConditioned2023a}. In parallel, many recent studies, inspired by knowledge distillation \cite{hintonDistillingKnowledgeNeural2015}, have shifted their focus to feature-level reconstruction. Reverse Distillation (RD) \cite{dengAnomalyDetectionReverse2022} has become popular due to its efficient and simple structure, and numerous innovative improvements \cite{tienRevisitingReverseDistillation2023,tienRevisitingReverseDistillation2023,wangMiniMaxADLightweightAutoencoder2024, guoDinomalyLessMore2024} have been developed based on it. For example, \cite{guoDinomalyLessMore2024} explored extending RD to Vision Transformers \cite{dosovitskiyImageWorth16x162021}, proposing key elements such as loose reconstruction, noise bottlenecks, and linear attention \cite{katharopoulosTransformersAreRNNs2020}, all of which play a significant role.
\\
\textbf{Synthetic Anomaly.}
Methods that synthetic anomalies provide valuable supervisory signals by generating synthetic anomalies. For example, CutPaste \cite{liCutPasteSelfsupervisedLearning2021b} creates anomalies through random cutting and pasting, while DRAEM \cite{zavrtanikDRAEMDiscriminativelyTrained2021} applies Perlin noise and an auxiliary texture dataset to generate anomalies. DeSTSeg \cite{zhangDeSTSegSegmentationGuided2022} uses a similar anomaly synthetic strategy in conjunction with feature reconstruction methods. Additionally, SimpleNet introduces Gaussian noise in the feature space to simulate anomalies. However, the generalization of these synthetic anomaly methods is limited. 
\\
\textbf{Zero/few-shot.}
The development of visual-language models (VLMs) has significantly advanced zero/few-shot anomaly detection. For example, \cite{jeongWinCLIPZeroFewshot2023,zhouAnomalyCLIPObjectagnosticPrompt2023} achieved both zero-shot and few-shot anomaly detection using the VLM model CLIP. Additionally, researchers leveraged the characteristics of sensory anomaly detection---where normal image patches find many similar patches in other unlabeled images, while anomalous patches find few---and proposed a mutual scoring strategy for images \cite{liMuScZeroshotIndustrial2024,dammAnomalyDINOBoostingPatchbased2024a}. This approach supports few-shot anomaly detection without VLMs and can extend to zero-shot anomaly detection within batches.

\subsection{Learning with Noisy Data}
Learning with Noisy Labels (LNL) and Semi-Supervised Learning (SSL) have gained significant attention for their practical applications. Some studies \cite{veitLearningNoisyLargeScale2017,yaoInstancedependentLabelnoiseLearning2021,liuIdentifiabilityLabelNoise2023} suggest that label noise distribution is learnable, allowing labels to be restored by modeling this distribution. \cite{huangO2UNetSimpleNoisy2019} uses loss values from multiple training rounds as a basis for sample selection. Other researchers \cite{hanCoteachingRobustTraining2018,weiCombatingNoisyLabels2020} mitigate the impact of noise through information exchange and fusion between multiple homogeneous models. In SSL, UPS \cite{rizveDefensePseudoLabelingUncertaintyAware2021} applies MC-Dropout \cite{galDropoutBayesianApproximation2016} to estimate prediction uncertainty, which helps correct model output and select more reliable pseudo-labels.

However, adding noise in unsupervised sensory anomaly detection \cite{yangGeneralizedOutDistributionDetection2024} introduces new challenges, as there are no reliable samples for model calibration, and normal and anomalous samples in sensory AD are difficult to distinguish. Additionally, sensory contamination in reconstruction models can directly cause ``identical shortcuts'' \cite{youUnifiedModelMulticlass2022}, leading to substantial model contamination. To address these issues, many existing methods \cite{qiuLatentOutlierExposure2022,chenDeepOneclassClassification2022, yoonSelfsuperviseRefineRepeat2022a} continue to apply semantic AD approaches to sensory AD, leading to a significant performance gap when compared to state-of-the-art sensory AD methods that do not incorporate Noisy. Recently, SoftPatch \cite{jiangSoftPatchUnsupervisedAnomaly2022}, based on PatchCore \cite{rothTotalRecallIndustrial2022}, proposed a patch-level filtering strategy that first filters out some anomalous patches, then stores them in a weighted Memory Bank, where each patch's anomaly score is calculated based on the distance to its nearest patch during inference. Although this method narrows the performance gap, its scalability and effectiveness remain limited.

\section{Method}
\label{sec:method}

Figure~\ref{fig:framework} shows the overall framework of the SSFilter. In the following sections, we introduce the components of SSFilter. Section \ref{ela} presents our improved baseline model. Section \ref{pcs} describes an efficient batch filter based on patch-wise comparative scoring. In Section \ref{aou}, we introduce a fine-grained uncertainty estimator that detects subtle anomalies in noisy images by leveraging uncertainty awareness. As discussed in Section \ref{slf}, we integrate these two filtering mechanisms to create a robust sample-level noise filter. Despite these efforts, indistinguishable noisy data can still interfere with the model, causing it to learn abnormal patterns. To address this, Section \ref{rtht} proposes a integrity enhancement scheme that recovers normal features from realistically generated synthetic anomalies to guide the model and counter overlooked anomalies. Finally, in Section \ref{eftd}, we present a method for scaling a well-trained model to dataset-level filtering.

\subsection{Dinomaly with Enhancing Linear Attention}
\label{ela}
Our baseline model incorporates Dinomaly \cite{guoDinomalyLessMore2024}, a reconstruction model with strong performance, using DINOv2 \cite{darcetVisionTransformersNeed2023, oquabDINOv2LearningRobust2024a} as the encoder. DINOv2 provides powerful patch feature representation capabilities, which are essential for the comparative scoring filter. The original Dinomaly extends the RD framework to Vision Transformers, featuring an encoder, a noise bottleneck, and a linear attention \cite{katharopoulosTransformersAreRNNs2020} decoder. Their experiments show that linear attention's ``non-focusing ability'' enhances the reconstruction model's performance. However, a prevailing view is that linear attention lacks expressive capacity \cite{choromanskiRethinkingAttentionPerformers2022, qinCosFormerRethinkingSoftmax2022}, which limits decoder effectiveness. To address this, we incorporate Mamba-Like Linear Attention (MLLA) \cite{hanDemystifyMambaVision2024}, an enhanced form of linear attention inspired by Mamba \cite{guMambaLinearTimeSequence2024}. MLLA integrates key elements of Mamba into linear attention, yielding strong performance improvements.

\subsection{Patch Comparison Scoring}
\label{pcs}
It is widely accepted that anomalies are outliers, characterized by their specificity and rarity. The concept of mutual scoring between image patches within a batch \cite{liMuScZeroshotIndustrial2024} builds on this notion. 
In essence, it assumes that normal image patches can find several similar patches within a batch, while anomalous patches are difficult to match. 
Its performance depends on two core components: the patch feature descriptor and the mutual scoring strategy.
\\
\textbf{Patch Feature Descriptor.}
We leverage DINOv2 as a powerful feature extractor, with the output of each ViT block naturally serving as a Patch Feature Descriptor. We denote $\phi_{i,j} = \phi_{j}(x_{i}) \in \mathbb{R}^{N \times C}$ as the feature embedding of image $x_i$ at the $j$-th layer of the pretrained network $\phi$, where $N$ is the number of patches and $C$ is the feature dimension.

Following the guidance of previous studies \cite{defardPaDiMPatchDistribution2021, rothTotalRecallIndustrial2022, liPromptADLearningPrompts2024}, selecting middle-level features is crucial, as deeper feature embeddings tend to lose the local features that are critical for describing a relatively independent patch. To combine features from different levels, $\phi_{i,j}$, we concatenate them along the feature dimension. In addition to single-layer descriptions, we also concatenate averaged features from multiple levels to obtain a more comprehensive representation. In summary, features from different levels combine complementary information, mitigating the limitations of single-layer features. We define this process as:
\begin{equation}
  \label{eq:l1l2}
  z^i=f_{cat}(\{\phi_{i,l}|l \in L_1\}, mean(\{\phi_{i,l} | l \in L_2\})).
\end{equation}
\\
\textbf{Mutual Scoring Strategy.}
Most advanced batch zero-shot anomaly detection methods require extended inference times. For both MuSc \cite{liMuScZeroshotIndustrial2024} and AnomalyDINO \cite{dammAnomalyDINOBoostingPatchbased2024a}, the inference time for a single image on an RTX 3090 GPU exceeds 100ms, making it challenging to use them directly as batch-level filters during training. To address this, we simplify the process in the following steps: first, we use a single matrix multiplication operation to compute the mutual similarity of all patches within the mini-batch. Second, for each patch, we calculate the average distance of the top $0.1\%$ most similar patches, which serves as the final normalcy score for that patch. Finally, similar to the operation in many UAD algorithms that compute anomaly values from anomaly maps \cite{dammAnomalyDINOBoostingPatchbased2024a, guoDinomalyLessMore2024, guoReContrastDomainspecificAnomaly2023}, we compute the average anomaly score of the top $1\%$ most anomalous regions to obtain the image-level anomaly score $a_i$. The core pseudocode of the mutual scoring method is presented in Algorithm \ref{alg:scoring}.

\begin{algorithm}[tb]
  \caption{PyTorch-style pseudocode for patch comparison scoring.}
  \label{alg:scoring}
  \definecolor{codeblue}{rgb}{0.25,0.5,0.5}
  \lstset{
    basicstyle=\fontsize{7.2pt}{7.2pt}\ttfamily\bfseries,
    commentstyle=\fontsize{7.2pt}{7.2pt}\color{codeblue},
    keywordstyle=\fontsize{7.2pt}{7.2pt},
  }
  \begin{lstlisting}[language=python]
  # x     - minibatch of images [B, C, H ,W]
  # N     - total patch count, value BxHxW
  # D     - feature dimensions of patch descriptors
  # 
  # mean  - mean value function
  # topk  - function that takes the k largest values.
  
  # compute patch feature descriptor
  z = f(x)   # [N, D]
  # normalized feature embedding
  z_norm =  z / z.norm(dim=1)
  # calculate the similarity matrix of patches
  sims = z_norm @ z_norm.T   # [N, N]
  # fill in the diagonal
  fill_diagonal_blocks(sims, HxW, 0)
  # calculate the patch-level anomaly score  # [N,]
  p_s = 1 - mean(topk(sims, k=N * 0.001, dim=1), dim=1)
  p_s = p_s.view(B,H*W)   # [B, H*W]
  # calculate the image-level anomaly score  # [B,]
  i_s = mean(topk(p_s, k=H * W * 0.01, dim=1), dim=1)
\end{lstlisting}
\end{algorithm}

\subsection{Awareness of Uncertainty}
\label{aou}
Since the patch scoring mechanism is an untrained zero-shot method, its performance is inherently constrained. To mitigate over-reliance on this approach, we propose a fine-grained, uncertainty-aware method. Specifically, we leverage the uncertainty caused by anomalous regions to refine the patch scoring mechanism. Furthermore, we extract real anomalous content from cumulative errors to generate realistic synthetic anomalies.
\\
\textbf{Uncertainty Estimate.}
Formally, let $F_i,\hat{F}_i \in \mathbb{R}^{C \times H \times W}$  represent the feature maps output by the encoder and decoder for the \( i \)-th image, where \( C \), \( H \), and \( W \) denote the channel count, height, and width of the output feature map, respectively. To estimate fine-grained uncertainty, the stacked decoder does not directly generate noise to avoid significant contamination of the alignment feature maps. Instead, we apply dropout \( K \) times at the noise bottleneck to generate \( K \) masked models. The anomalous images generated during each forward pass are as follows:
\begin{equation}
  M_i^k = \mathcal{D}_M(F_{i}, \hat{F}_{i,W^k}).
\end{equation}
where \( W^k \) denotes the weights of the \( k \)-th masked model, and \( \mathcal{D}_M \) represents a feature alignment function that generates anomaly maps. The anomaly score is computed as the mean of the maximum value regions:
\begin{equation}
  s_i^k = \mathop{max}\limits_{0.01}(M_i^k).
\end{equation}
The prediction uncertainty \(u_i\) of \(x_i\) is given by the standard deviation of \(s_i \in \mathbb{R}^k\)
\\
\textbf{Anomalous Material Extraction.}
Anomalous regions exhibit both high regression error and high prediction uncertainty, allowing us to extract anomalous material from anomaly samples. By combining stable foreground estimation, we can synthesize realistic anomalous samples. We accumulate $K$ anomalous images as follows:
\begin{equation}
  M_i^{acc} = \sum_{k=1}^{K} M_i^k.
\end{equation}
This anomaly map accumulates both regression error and prediction uncertainty. Using a threshold, we crop the most anomalous regions as anomaly material, as shown in Figure~\ref{fig:framework}.
\subsection{Synergy Scoring Filtering}
\label{slf}
Since reliable uncertainty estimation relies on good reconstruction ability, we divide the training process into two main phases: the cold-start phase and the regular training phase. By default, the cold-start phase corresponds to the first 1,000 iterations of training.
\\
\textbf{Cold Start Phase.}
Since anomalous samples are present during training in the FUAD setting, our goal is to minimize the impact of these anomalies to improve performance. In the cold-start phase, the results from patch comparison scoring will serve as our primary reference. Since anomaly levels vary across different sample categories and sizes, applying a fixed threshold for filtering becomes challenging. Therefore, we employ a general filtering strategy, ranking the samples in a mini-batch and selecting the top half with high confidence for training. Formally, given a sorted set of anomaly scores \( A^{\prime} = \{a^{\prime}_1, a^{\prime}_2, \dots, a^{\prime}_n\} \), where \( a^{\prime}_1 \leq a^{\prime}_2 \leq \dots \leq a^{\prime}_n \), we split the data into two subsets \( A^{\prime}_1 \) and \( A^{\prime}_2 \) as follows:
\begin{equation}
  A^{\prime}_1 = \{a^{\prime}_1, a^{\prime}_2, \dots, a^{\prime}_m\}, \quad A^{\prime}_2 = \{a^{\prime}_{m+1}, a^{\prime}_{m+2}, \dots, a^{\prime}_n\}.
\end{equation}
where the split point \( m\) is \( \frac{n}{2} \). The corresponding uncertainty is denoted as:
\begin{equation}
  U^{\prime}_1 = \{u^{\prime}_1, u^{\prime}_2, \dots, u^{\prime}_m\}, \quad U^{\prime}_2 = \{u^{\prime}_{m+1}, u^{\prime}_{m+2}, \dots, u^{\prime}_n\}.
\end{equation}
At this stage, we define the samples corresponding to the elements of \( U^{\prime}_1 \) as a stable set of normal samples, where the mean and variance of \( U^{\prime}_1 \) are denoted by \( \mu_1 \) and \( \sigma_1 \), respectively. These statistics represent the uncertainty anchor points for normal samples and uncertainty fluctuations in normal samples. Let \( \boldsymbol{g} = \{ g^{(1)}, \dots, g^{(n)} \} \subseteq \{ 0, 1 \}^n \) be a binary vector, where \( g^{(i)} \) indicates whether sample \( i \) is selected as a training sample. The formula for calculating this vector is as follows:
\begin{equation}
  \label{eq:select}
  g^{(i)}=\mathbb{I}\left[a_i\leq a_m^{\prime} \right]+\mathbb{I}\left[a_i \geq a_{m+1}^{\prime} \right]\mathbb{I}\left[u_i \leq \mu_1\right].
\end{equation}
The first term indicates that half of the samples with lower anomaly scores will be used as training samples, while the second term introduces a slight intervention of uncertainty by using the normal anchor point \( \mu_1 \) to recall some samples with higher scores as training samples. The set of samples selected for training within the batch is denoted as:
\begin{equation}
  \mathcal{X}_g=\bigcup_{g^{(i)}=1}x_i.
\end{equation}
\\
\textbf{Regular Training Phase.}
To mitigate this one-sided strong dependency, we integrate uncertainty at this stage. Specifically, for the anomaly estimation scores prior to ranking, we combine them with uncertainty, calculated as follows:
\begin{equation}
  A \leftarrow \frac{1}{2}\left(norm(A)+norm(U)\right).
\end{equation}
where the norm defaults to min-max normalization. It is important to note that \( \mu_1 \) and \( \sigma_1 \) are still calculated from the pre-update \( A^{\prime} \) and \( U^{\prime} \) to prevent early leakage, while the remaining components are consistent with the cold-start phase.

\subsection{Restoring the Hidden Truth}
\label{rtht}
\begin{figure}[t]
  \centering
  \includegraphics[width=0.8\linewidth]{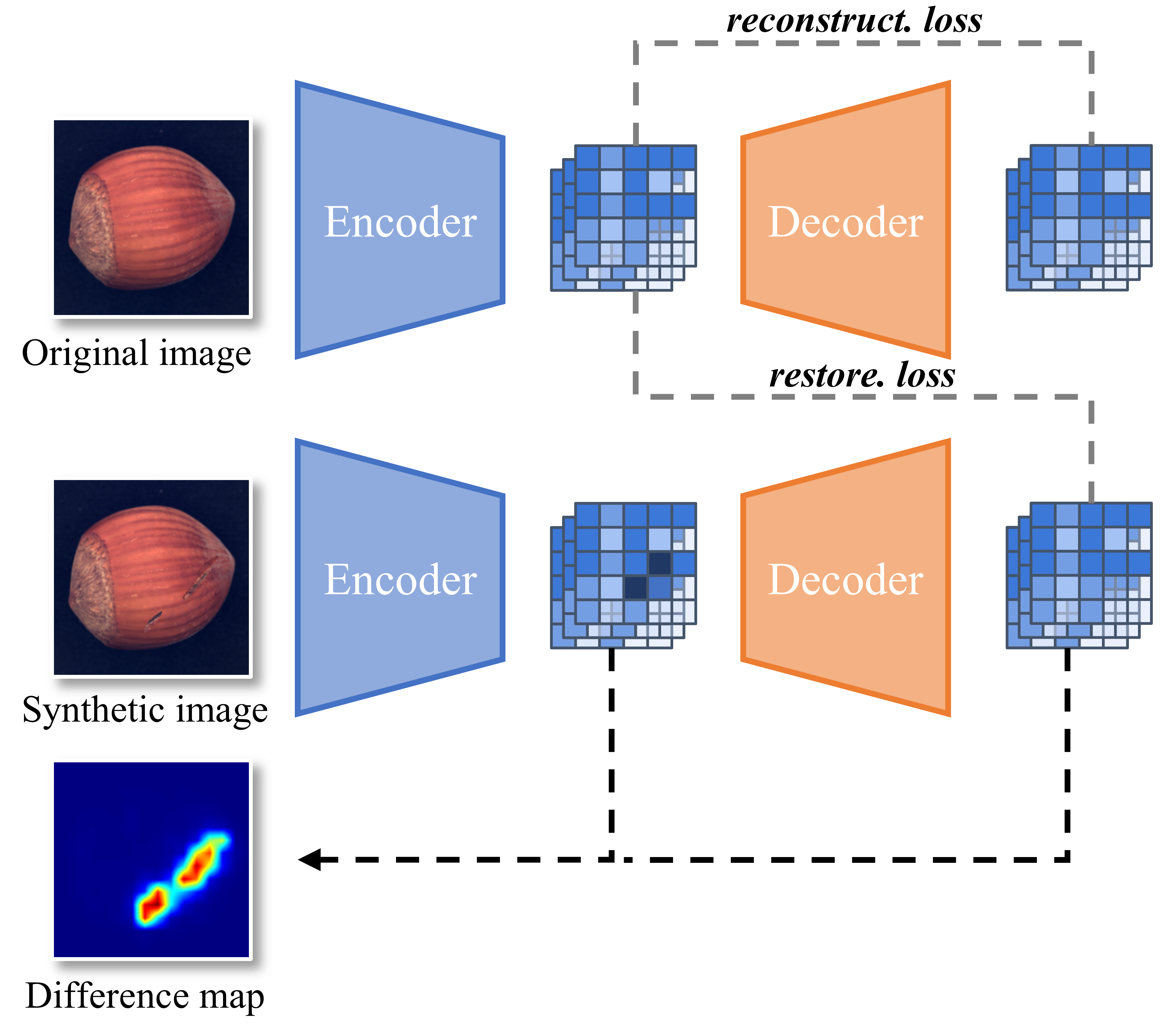}
  \caption{The model restores the original appearance of the object from the real synthetic anomalies.}
  \label{fig:restore}
\end{figure}

We follow best practices in FUAD, assuming the noise rate is unknown in advance. To maximize the supervisory signals provided by potential anomaly samples in the training set, a non-destructive outlier exposure method is essential. This method must be capable of operating robustly, even when the noise rate is zero. We synthesize realistic anomalies using normal images and then restore their original appearance from the synthetic anomalies. This approach improves the model's ability to model normal images, which in turn aids in generating fine-grained segmentation maps for anomalous samples. An intuitive illustration of this process is shown in Figure~\ref{fig:restore}.
\\
\textbf{Anomalous Material Bank Construction.}
To uncover potential real anomalies, further filtering is required. Let \(\boldsymbol{h} = \{h^{(1)}, \dots, h^{(n)}\} \subseteq \{0, 1\}^n\) be a binary vector, where \(h^{(i)}\) indicates whether sample \(i\) is selected for constructing the anomaly bank. The calculation formula for this vector is as follows:
\begin{equation}
  \label{eq:aselect}
  h^{(i)}=\mathbb{I}\left[a^{(i)}\geq a_{m+1}\right]\mathbb{I}\left[u^{(i)}\geq \mu_1 + \tau \times \sigma_1  \right].
\end{equation}
where \(\tau\) serves as a hyperparameter for tuning outlier detection. A larger value of \(\tau\) indicates that a sample's uncertainty deviates more from the anchor by multiples of normal fluctuation, thereby being considered a high-confidence anomaly. For each selected anomaly in the minibatch, we construct an anomaly material bank using the method described in Section \ref{aou}:
\begin{equation}
  \label{eq:bank}
  \mathcal{M}=\bigcup_{h^{(i)}=1}\{x_i \in \mathcal{X}_h | M_i^{acc} \geq T_p \}.
\end{equation}
where $T_p$ is a threshold for selecting the largest region.
\\
\textbf{Synthetic Anomalies and Optimization Goals.}
To synthesize real anomalies, we used a PCA-based foreground estimator. By applying PCA to the features generated by the encoder, we obtain a stable object mask. Next, we randomly sample from \(\mathcal{X}_g\). For each sampled image \(x_i^* \in \mathcal{X}_g\), we select several  anomalous elements from \(\mathcal{M}\), apply random rotations, and paste them onto the foreground region, resulting in the synthetic anomaly image \(\widetilde{x}_i^* \in \widetilde{\mathcal{X}_g}\). Finally, our optimization objective includes both a reconstruction loss and a restoration loss, defined as:
\begin{equation}
  \begin{aligned}
    \mathcal{L}_{rec.} & = \frac{1}{n_1+n_2} \sum_{i=0}^{n_1} \mathcal{D}\left(F\left(x_i\right), \hat{F}\left(x_i\right)\right),
  \end{aligned}
\end{equation}

\begin{equation}
  \begin{aligned}
    \mathcal{L}_{res.} & = \frac{1}{n_1+n_2} \sum_{i=0}^{n_2} \mathcal{D}\left(F\left(x_i\right), \hat{F}\left(\widetilde{x}_i^*\right)\right),
  \end{aligned}
\end{equation}

\begin{equation}
  \begin{aligned}
    \mathcal{L} = \mathcal{L}_{rec.} + \mathcal{L}_{res.}.
  \end{aligned}
\end{equation}
where $n_1=|\mathcal{X}_g|$, $n_2=|\mathcal{\widetilde{X}}_g|$ and $\mathcal{D}$ is the hard-mining cosine distance loss function \cite{guoReContrastDomainspecificAnomaly2023}.

\begin{table*}[t]
  \centering
  \caption{Fully unsupervised anomaly detection and segmentation results with I-AUROC/P-AUPRO(\%) on Real-IAD. Methods marked with a superscript $\dagger$ indicate that the training set was filtered using the well-trained SSFilter$_{448392}$ model. Subscripts in the table visually depict changes relative to the original experimental data.}
  \begin{tabular}{l|cccccc}
    \toprule
    Methods                                       & $\alpha = 0.0$                                                      & $\alpha = 0.1$                                                      & $\alpha = 0.2$                                                      & $\alpha = 0.4$                                                      & Mean                                                                \\
    \midrule
    RD\cite{dengAnomalyDetectionReverse2022}                                            & 91.8/94.1                                                           & 90.2/94.7                                                           & 89.4/94.8                                                           & 87.8/\underline{94.3}                                               & 89.8/94.5                                                           \\
    Dinomaly\cite{guoDinomalyLessMore2024}                                      & \textbf{93.9}/\textbf{95.5}                                         & 91.6/\textbf{96.1}                                                  & 90.0/\textbf{95.5}                                                  & 87.4/\textbf{94.8}                                                  & 90.7/\textbf{95.5}                                                  \\
    PatchCore\cite{rothTotalRecallIndustrial2022}                                     & 91.9/89.8                                                           & 90.2/89.0                                                           & 88.7/88.3                                                           & 86.3/86.4                                                           & 89.3/88.4                                                           \\
    SoftPatch\cite{jiangSoftPatchUnsupervisedAnomaly2022}                                     & 91.4/90.7                                                           & 90.7/91.0                                                           & 90.1/90.8                                                           & 88.5/89.8                                                           & 90.2/90.6                                                           \\
    DeSTSeg\cite{zhangDeSTSegSegmentationGuided2022}                                       & 91.1/94.5                                                           & 88.9/93.3                                                           & 86.3/91.5                                                           & 82.1/89.8                                                           & 87.1/92.3                                                           \\

    \rowcolor{bluee} \textbf{SSFilter$_{256224}$} & 92.4/90.7                                                           & \underline{92.3}/91.6                                               & \underline{92.1}/91.5                                               & \underline{90.9}/91.1                                               & \underline{91.9}/91.2                                               \\
    \rowcolor{bluee} \textbf{SSFilter$_{448392}$} & \underline{92.5}/\underline{94.9}                                   & \textbf{92.7}/\underline{95.2}                                      & \textbf{92.6}/\underline{95.1}                                      & \textbf{91.8}/\textbf{94.8}                                         & \textbf{92.4}/\underline{95.0}                                      \\
    \midrule
    \rowcolor{pink!8} RD$^\dagger$                & 89.2\textcolor{down}{{$_{-2.6}$}}/92.8\textcolor{down}{{$_{-1.3}$}} & 90.3\textcolor{upup}{{$_{+0.1}$}}/94.1\textcolor{down}{{$_{-0.6}$}} & 90.1\textcolor{upup}{{$_{+0.7}$}}/94.4\textcolor{down}{{$_{-0.4}$}} & 89.6\textcolor{upup}{{$_{+1.8}$}}/94.4\textcolor{upup}{{$_{+0.1}$}} & 89.8\textcolor{upup}{{$_{+0.0}$}}/93.9\textcolor{down}{{$_{-0.6}$}} \\
    \rowcolor{pink!8} Dinomaly$^\dagger$          & 92.5\textcolor{down}{{$_{-1.4}$}}/95.2\textcolor{down}{{$_{-0.3}$}} & 92.4\textcolor{upup}{{$_{+0.8}$}}/95.5\textcolor{down}{{$_{-0.6}$}} & 91.9\textcolor{upup}{{$_{+1.9}$}}/95.7\textcolor{upup}{{$_{+0.2}$}} & 90.5\textcolor{upup}{{$_{+3.1}$}}/95.3\textcolor{upup}{{$_{+0.5}$}} & 91.8\textcolor{upup}{{$_{+1.1}$}}/95.4\textcolor{down}{{$_{-0.1}$}} \\
    \rowcolor{pink!8} PatchCore$^\dagger$         & 90.9\textcolor{down}{{$_{-1.0}$}}/89.2\textcolor{down}{{$_{-0.6}$}} & 90.5\textcolor{upup}{{$_{+0.3}$}}/89.5\textcolor{upup}{{$_{+0.5}$}} & 90.1\textcolor{upup}{{$_{+1.4}$}}/89.4\textcolor{upup}{{$_{+1.1}$}} & 88.8\textcolor{upup}{{$_{+2.5}$}}/88.4\textcolor{upup}{{$_{+2.0}$}} & 90.1\textcolor{upup}{{$_{+0.8}$}}/89.1\textcolor{upup}{{$_{+0.7}$}} \\
    \rowcolor{pink!8} SoftPatch$^\dagger$         & 89.9\textcolor{down}{{$_{-1.5}$}}/89.4\textcolor{down}{{$_{-1.3}$}} & 89.8\textcolor{down}{{$_{-0.9}$}}/89.9\textcolor{down}{{$_{-1.1}$}} & 89.7\textcolor{down}{{$_{-0.4}$}}/90.3\textcolor{down}{{$_{-0.5}$}} & 89.3\textcolor{upup}{{$_{+0.8}$}}/90.2\textcolor{upup}{{$_{+0.4}$}} & 89.7\textcolor{down}{{$_{-0.5}$}}/90.0\textcolor{down}{{$_{-0.6}$}} \\
    \rowcolor{pink!8} DeSTSeg$^\dagger$           & 89.9\textcolor{down}{{$_{-1.2}$}}/94.8\textcolor{upup}{{$_{+0.3}$}} & 90.3\textcolor{upup}{{$_{+1.4}$}}/95.0\textcolor{upup}{{$_{+1.7}$}} & 90.7\textcolor{upup}{{$_{+4.4}$}}/94.7\textcolor{upup}{{$_{+3.2}$}} & 89.3\textcolor{upup}{{$_{+7.2}$}}/93.0\textcolor{upup}{{$_{+3.2}$}} & 90.1\textcolor{upup}{{$_{+3.0}$}}/94.4\textcolor{upup}{{$_{+2.1}$}} \\
    \bottomrule
  \end{tabular}
  \label{tab:comparison}

\end{table*}

\subsection{Extended Filtering to Datasets}
\label{eftd}
The filtering performance of SSFilter improves continuously during training until it reaches a bottleneck. A well-trained SSFilter acts as an effective sample-level anomaly filter on its own, making it suitable for filtering noisy datasets to produce cleaner datasets. This feature significantly broadens methodological options in practical applications, enabling users to choose the solution best suited to their specific scenario. Specifically, for a trained SSFilter, we perform random iterations over  \(\kappa_1\) epochs and classify a sample as normal if the filter identifies it as normal more than  \(\kappa_2\) times. By default, \(\kappa_1\) is set to 20, and \(\kappa_2\) is set to 5.

\section{Experiments}
\label{sec:exp}
\subsection{Experimental Settings}
\noindent
\textbf{Datasets.}
\textbf{Real-IAD} \cite{wangRealIADRealworldMultiview2024} is a large-scale, real-world multi-view dataset designed for industrial anomaly detection. It includes 30 distinct object categories, with 99,721 normal images and 51,329 abnormal images. A sufficiently large dataset enabled the creation of the first comprehensive unsupervised anomaly detection benchmark, which defines four noise levels: \{0.0, 0.1, 0.2, 0.4\}.
\textbf{MVTec AD} \cite{bergmannMVTecADComprehensive2019} is a well-known dataset for industrial anomaly detection, containing more than 5,000 high-resolution images across fifteen object and texture categories. The limited number of abnormal samples in MVTecAD makes it challenging to construct a reliable FUAD benchmark \cite{wangRealIADRealworldMultiview2024}, as a reduction in abnormal test samples compromises the credibility of the evaluation. To comprehensively evaluate the performance of our model across datasets of varying scales, we followed the setup of SoftPatch and reported the results of different models under a noise level of 0.1.
\\
\textbf{Implementation Details.}
By default, the ViT-Small/14 encoder (patch size = 14) is used, pretrained with DINOv2-R \cite{darcetVisionTransformersNeed2023}. The Noisy Bottleneck has a default dropout rate of 0.2, and the MLP ratio for the MLLA block in the decoder is set to 1.0. The default setting for the number of noise occurrences, K, used to estimate uncertainty is 10. In Equation \ref{eq:l1l2}, \(L_1\) is set to \{0, 3\}, and \(L_2\) to \{0, 3, 7\}, summarizing information from different layers. Additionally, in Equation \ref{eq:aselect}, \(\tau\) is set to 8. For \(T_p\) in Equation \ref{eq:bank}, the default is the minimum value between 80\% of its maximum value and its 99th percentile. In the comparison experiments, unless otherwise specified, the input image size is set to \(448^2\), followed by a central crop to \(392^2\) to ensure that the feature map (\(28^2\)) is sufficiently large for anomaly localization. In all ablation study, for efficiency, the input image size is set to \(256^2\) and then center-cropped to \(224^2\), resulting in a decrease in segmentation performance. The StableAdamW optimizer \cite{wortsmanStableLowprecisionTraining2023} with AMSGrad \cite{reddiConvergenceAdam2019} is used, with a learning rate of \(2 \times 10^{-3}\) and a decay rate of \(1 \times 10^{-4}\). The model was trained for 5,000 iterations on the Real-IAD dataset and 1,000 iterations on the MVTecAD dataset, using an NVIDIA RTX 3090 GPU (24GB) with a batch size of 32.

\begin{table}[t]
\centering
\begin{tabular}{cccc}
\toprule
\multicolumn{1}{c}{Setting} & Methods &  I-AUROC & P-AUPRO \\ \midrule
\multirow{6}{*}{No overlap} & RD & 97.3 & 92.6 \\ 
                        & SoftPatch & 98.5 & 92.9 \\ 
                        & PatchCore & 98.2 & 91.7 \\ 
                        & Dinomaly & \underline{98.9} & 90.4 \\ 
                        & SSFilter$_{256224}$ & 98.6 & \underline{93.6} \\ 
                        & SSFilter$_{448392}$ & \textbf{99.0} & \textbf{95.2} \\ 
                        \midrule
\multirow{6}{*}{Overlap} & RD & 85.4 & 91.8 \\ 
                            & SoftPatch & \underline{98.4} & 89.5 \\ 
                            & PatchCore & 68.4 & 63.9 \\ 
                            & Dinomaly & 93.9 & 90.2 \\ 
                            & SSFilter$_{256224}$ & 98.3 & \underline{92.5} \\ 
                            & SSFilter$_{448392}$  & \textbf{98.9} & \textbf{94.1} \\ 
                            \bottomrule
\end{tabular}
\caption{Fully unsupervised anomaly detection and segmentation results with I-AUROC and P-AUPRO on MVTecAD-noise-0.1. Overlap means the injected anomalous images are included in the test set.}
\label{tab:mvtec}
\end{table}
In the comparative experiments, we adopted default settings for unspecified cases, though certain scenarios necessitate further elaboration. For the Real-IAD dataset, its substantially larger scale compared to prior datasets prompted a reduction in RD's training epochs from 200 to 50 to minimize computational overhead and mitigate overfitting. Regarding PatchCore and SoftPatch, the extensive training data imposed significant memory constraints, elevating inference costs; thus, we lowered the core set downsampling rate from 0.1 to 0.01. In DeSTSeg experiments, we adhered to the authors' protocol, employing non-rotated configurations across all categories. All experiments on the MVTecAD dataset were performed using the default configuration.
\\
\textbf{Metrics.} Image-level anomaly detection performance is quantified using the Area Under the Receiver Operator Curve (I-AUROC). For anomaly localization, we utilize Area Under the Per-Region Overlap (P-AUPRO). P-AUPRO is designed to treat anomalous regions of various sizes equally, thus providing uniform sensitivity across different anomaly scales.

\begin{table*}[htbp]
  \centering
  \caption{Complexity compared with the baseline method. Frames Per Second (FPS) are measured on an NVIDIA RTX 3090 with a batch size of 16. mAD represents the average of the anomaly detection metric (I-AUROC) and the anomaly segmentation metric (P-AUPRO) across the four noise-level benchmarks. Time is the average training time for a single category.}
  \begin{tabular}{ccccccc}
      \toprule
      Method              & Resolution        & Backbone & Params(M)         & FLOPs(G)                 & FPS             & mAD                            \\
      \midrule
      Dinomaly\cite{guoDinomalyLessMore2024}            & R$448^2$-C$392^2$ & ViT-B    & \underline{148.0} & 114.9             & 80              & 90.7/\textbf{95.5}             \\
      SSFilter$_{256224}$ & R$256^2$-C$224^2$ & ViT-S    & \textbf{31.7}     & \textbf{7.8}    & \textbf{638}    & \underline{91.9}/91.2          \\
      SSFilter$_{448392}$ & R$448^2$-C$392^2$ & ViT-S    & \textbf{31.7}     & \underline{26.8}           & \underline{202} & \textbf{92.4}/\underline{95.0} \\
      \bottomrule
  \end{tabular}
  \label{tab:effectiveness}
\end{table*}

\subsection{Fully Unsupervised Anomaly Detection Experimental}
Table \ref{tab:comparison} presents the experimental results of various state-of-the-art (SOTA) models on the fully unsupervised Real-IAD benchmark. As shown in the table, noise addition in our comparative methods significantly impairs their performance, with degradation accelerating as the noise ratio increases. SoftPatch, which employs a patch-filtering mechanism designed for FUAD, maintains an advantage, though it still experiences a substantial performance drop. The proposed SSFilter performs slightly worse than its baseline model, Dinomaly, in noiseless settings. This performance gap is partly due to the filtering mechanism, which discards some normal samples, and partly due to adjustments made to improve training efficiency. Specifically, we configured the backbone as ViT-S instead of ViT-B, the default in Dinomaly, and reduced the MLP ratio from 4.0 to 1.0. These changes reduced the training time of SSFilter$_{256224}$ to approximately match that of standard Dinomaly. While this setup enhanced training and inference speed, it also reduced representation capability. Table \ref{tab:effectiveness} presents a detailed comparison of complexity, while Table \ref{tab:ab_key} compares performance at the same complexity level as the baseline method (i.e., the first row of Table \ref{tab:ab_key}). 

In addition, SoftPatch, based on PatchCore, also experienced performance degradation compared to its baseline in noise-free settings due to the discard mechanism, underscoring the importance of data-efficient models. At low noise levels (0.1 and 0.2), SSFilter avoids a steep performance drop; in SSFilter$_{448392}$, performance under low noise levels even surpasses that in noise-free settings. As noise levels increase, SSFilter's advantage becomes more pronounced, maintaining robust performance even at a noise setting of 0.4. For SSFilter$_{256224}$, the original feature map size is limited to $16^2$, meaning that the interpolated high-resolution feature map provides only coarse anomaly localization, resulting in suboptimal localization performance. Nonetheless, its anomaly detection performance remains comparable to that of SSFilter${_{448392}}$.

\begin{figure}[t]
  \centering
  \includegraphics[width=0.9\linewidth]{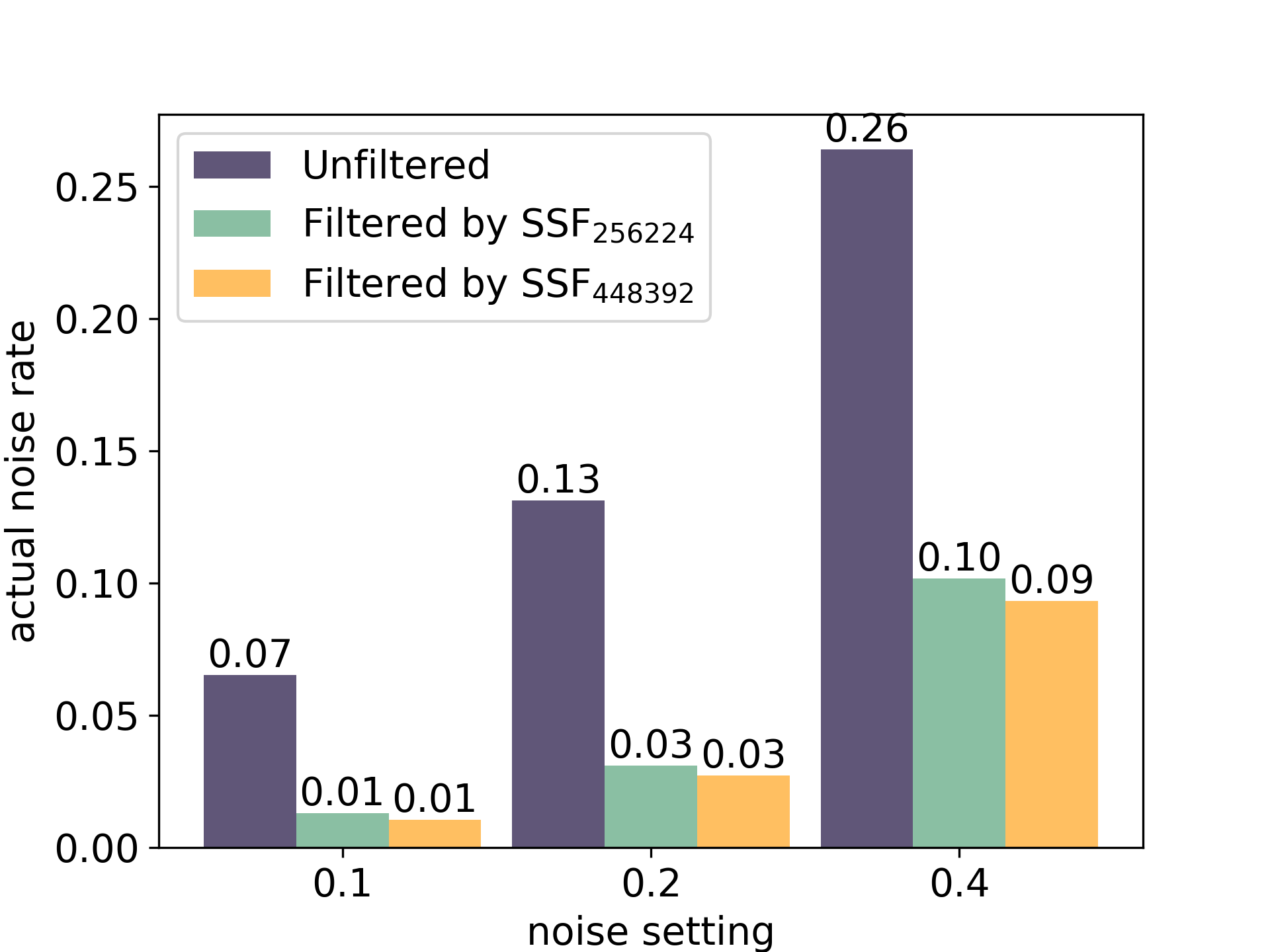}
  \caption{Quantitative results of dataset-level noise filtering are presented.}
  \label{fig:nr}
\end{figure}

\begin{figure}[t]
  \centering
  \includegraphics[width=0.9\linewidth]{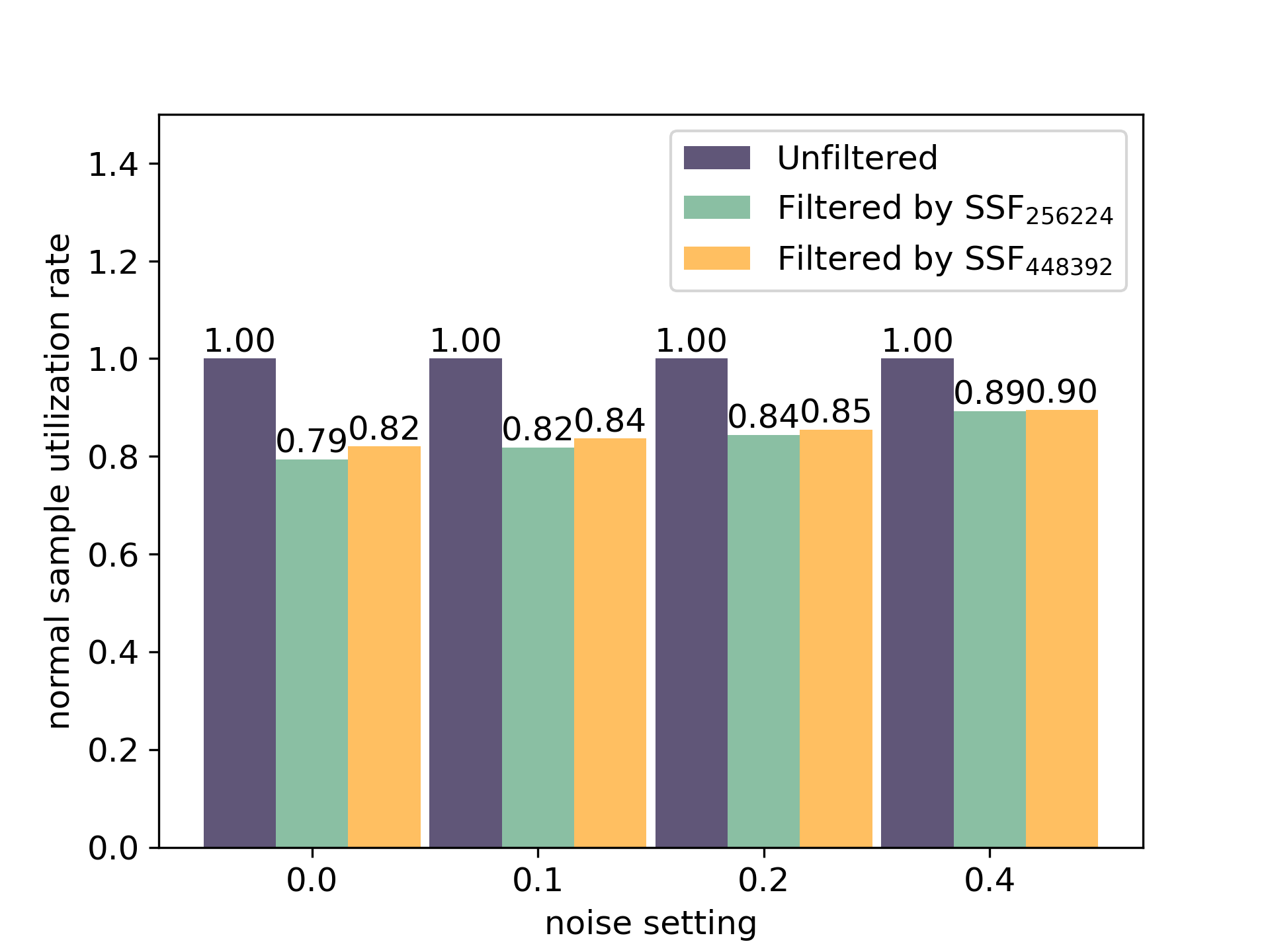}
  \caption{Quantitative results of the utilization rate of normal samples after dataset-level filtering.}
  \label{fig:ut}
\end{figure}

For dataset-level filtering, our practical tests indicate that the actual measured noise rates for the four noise settings in Real-IAD, \{0.0, 0.1, 0.2, 0.4\}, are \{0.0, 0.07, 0.13, 0.26\}. We calculated the noise rate as the ratio of anomaly samples to the total number of samples in the dataset. Figure~\ref{fig:nr} and \ref{fig:ut} display the noise rate and the normal sample utilization rate of the filtered dataset in comparison to the original dataset, demonstrating the dataset-level filtering capability of our method. Figure~\ref{fig:ut} indicates that higher noise ratio in the dataset facilitate the model's ability to learn discriminative information, improving normal sample utilization. Results in Table \ref{tab:comparison} further show that our method can extend to other approaches. In a noise-free environment, all models experienced a certain degree of performance degradation due to the incompleteness of the training set, reflecting the models' data efficiency. PatchCore, known for its robust few-shot performance, exhibited the smallest decline. In noisy environments, SoftPatch, specifically designed for FUAD, did not benefit from filtered dataset. However, when noise setting reached 0.4, SoftPatch's filtering strategy struggled to handle the high noise level, which ultimately resulted in improved performance on the filtered dataset. In most cases, our dataset-level filtering enhanced anomaly detection performance across methods, yielding excellent overall results, though with slight declines in some anomaly localization metrics. 

To evaluate the performance of SSFilter on small-scale data, we compare various SOTA methods on the MVTec AD dataset, as presented in Table \ref{tab:mvtec}. The results demonstrate that our method outperforms all others in both experimental settings. Under the No Overlap setting, our baseline method, Dinomaly, exhibits strong robustness; however, its anomaly segmentation performance remains notably inferior to our improved approach. SoftPatch, a method specifically designed for fully unsupervised scenarios, shows marginal improvements over PatchCore, though these enhancements remain limited. In the Overlap setting, RD, PatchCore, and Dinomaly all suffer significant performance degradation, whereas methods tailored for fully unsupervised scenarios retain robustness. Notably, our method consistently achieves superior results. For detailed experimental results for each category, please refer to the supplementary materials.

\subsection{Ablation Study}
\begin{table}[h]
  \caption{Ablation study on key designs, reporting I-AUROC/P-AUPRO (\%) on the Real-IAD dataset. \textbf{MLLA}: The decoder uses Mamba-Like Linear Attention. \textbf{PScoring}: Filtering is performed using the patch comparison scoring mechanism. \textbf{UScoreing}: Filtering is performed using uncertainty estimation methods. \textbf{URecall}: The uncertainty method participates in the recall of normal samples. \textbf{Restoring}: Restore synthesis anomalies.}
  \setlength\tabcolsep{3.0pt}
  \resizebox{1.0\linewidth}{!}{
    \begin{tabular}{ccccc|cc}
      \toprule
      \textbf{MLLA}          & \textbf{PScoring} & \textbf{UScoreing} & \textbf{URecall} & \textbf{Restoring} & $\alpha = 0.0$                    & $\alpha = 0.4$                 \\
      \toprule
      \xmarkg                & \xmarkg           & \xmarkg            & \xmarkg          & \xmarkg            & \underline{92.5}/\underline{91.3} & 86.9/90.8                      \\
      \hline
      \cmark                 & \xmarkg           & \xmarkg            & \xmarkg          & \xmarkg            & \textbf{93.0}/\textbf{91.6}       & 87.6/\textbf{91.3}             \\
      \xmarkg                & \cmark            & \xmarkg            & \xmarkg          & \xmarkg            & 90.8/90.0                         & 88.5/90.2                      \\
      \xmarkg                & \xmarkg           & \cmark             & \xmarkg          & \xmarkg            & 90.7/89.1                         & 88.9/90.6                      \\
      \hline
      \cmark                 & \cmark            & \xmarkg            & \xmarkg          & \xmarkg            & 91.5/90.3                         & 89.1/90.7                      \\
      \cmark                 & \cmark            & \xmarkg            & \xmarkg          & \cmark             & 92.2/90.3                         & 89.6/90.0                      \\
      \cmark                 & \cmark            & \xmarkg            & \cmark           & \cmark             & \underline{92.5}/90.5             & \underline{90.0}/90.7          \\
      \hline
      \rowcolor{bluee}\cmark & \cmark            & \cmark             & \cmark           & \cmark             & 92.4/90.7                         & \textbf{90.9}/\underline{91.1} \\
      \bottomrule
    \end{tabular}
  }
  \label{tab:ab_key}
\end{table}

\begin{table}[htbp]
  \centering
  \caption{Comparison of anomalous synthesis methods, presenting I-AUROC/P-AUPRO(\%) results on the Real-IAD dataset.}
  \begin{tabular}{l|ccc}
      \toprule
      methods               & $\alpha = 0.0$ & $\alpha = 0.4$ \\
      \toprule
      cutpaste\cite{liCutPasteSelfsupervisedLearning2021b}             & 92.0/90.4      & 90.0/90.6      \\
      draem\cite{zavrtanikDRAEMDiscriminativelyTrained2021}                & 92.1/90.4      & 89.9/90.5      \\
      \rowcolor{bluee}ours & 92.4/90.7      & 90.9/91.1      \\
      \bottomrule
  \end{tabular}
  \label{tab:ab_syn}
\end{table}

\noindent\textbf{Overall effectiveness comparison of the key elements proposed.}
We first examine the impact of the proposed key elements on model performance. As shown in Table~\ref{tab:ab_key}, MLLA directly enhances the model's overall performance. The patch comparison scoring and uncertainty estimation filtering methods independently improve UAD performance in noisy environments. However, because half of the samples in each batch are discarded in each iteration, these methods result in greater performance degradation in noise-free environments. When a single filtering strategy is applied, the introduction of MLLA still enhances overall model performance. Furthermore, incorporating the restoring strategy improves performance across varying noise ratios. URecall, defined as the second term in Equation \ref{eq:select}, successfully recovers a portion of normal samples from the discarded half, further improving model performance. Finally, by combining the synergistic scoring method outlined in Section \ref{slf}, the robust SSFilter is achieved, allowing the model to attain optimal performance.
\begin{table}[htbp]
    \centering
    \caption{Hyperparameter robustness experiment on the outlier screening hyperparameter $\tau$  in Equation~\ref{eq:select}, presenting I-AUROC/P-AUPRO(\%) results on the Real-IAD dataset.}
    \begin{tabular}{l|ccc}
        \toprule
        $\tau$            & $\alpha = 0.0$ & $\alpha = 0.4$ \\
        \toprule
        6                 & 92.4/90.7      & 90.7/90.9      \\
        \rowcolor{bluee}8 & 92.4/90.4      & 90.8/91.0      \\
        10                & 92.3/90.3      & 90.7/91.0      \\
        \bottomrule
    \end{tabular}
    \label{tab:ab_tau}
\end{table}
\begin{table}[htbp]
    \centering
    \caption{Hyperparameter robustness experiments for threshold $T_p$ on Real-IAD, presenting I-AUROC/P-AUPRO(\%) results on the Real-IAD dataset.}
    \begin{tabular}{c|cc}
        \toprule
        Percentage of maximum & \(\alpha=0.0\) & \(\alpha=0.4\) \\
        \toprule
        70\%                  & 92.1/90.0      & 90.7/90.7      \\
        \rowcolor{bluee}80\%  & 92.4/90.7      & 90.9/91.1      \\
        90\%                  & 92.3/90.6      & 90.8/91.1      \\
        \bottomrule
    \end{tabular}
    \label{tab:ab_Tp}
\end{table}
\begin{table}[htbp]
  \centering
  \setlength\tabcolsep{3.0pt}
  \caption{Ablation studies on patch descriptors.}
  \resizebox{0.6\linewidth}{!}{
  \begin{tabular}{cc|c}
    \toprule
    $L_1$                  & $L_3$     & Abnormal Recall \\
    \toprule
    \{0\}                  & \{-\}     & 0.7831          \\
    \{3\}                  & \{-\}     & 0.8864          \\
    \{7\}                  & \{-\}     & 0.8742          \\
    \{0,3\}                & \{-\}     & 0.8773          \\
    \{3,7\}                & \{-\}     & 0.8765          \\
    \{3\}                  & \{0,2,4\} & 0.8824          \\
    \{0,3\}                 & \{0,2,4\} & 0.8803         \\
    \{0\}                  & \{0,3,7\} & 0.8943          \\
    \hline
    \rowcolor{bluee}\{0,3\} & \{0,3,7\} & 0.9027         \\
    \bottomrule
  \end{tabular}
  }
  \label{tab:ab_pfd}
\end{table}
\\
\textbf{Ablation of the anomalous synthesis methods.}
For the two anomaly synthesis methods, CutPaste and DRAEM, we use the default configuration, generating two anomalous images per batch and applying a mask to ensure synthesis occurs in the foreground area. The results in Table \ref{tab:ab_syn} indicate that the proposed real anomaly synthesis strategy performs well, achieving optimal performance even at a noise rate of 0.0. Moreover, it demonstrates robustness in high-noise environments, highlighting the effective utilization of supervised signals in anomalous samples.
\\
\textbf{Ablation of the patch feature descriptor.}
In Table~\ref{tab:ab_pfd}, we compare the impact of different feature levels on the patch feature descriptor. This experiment uses both the training and testing sets of MVTecAD, with the recall rate of abnormal samples among discarded samples in each batch as the evaluation metric. Given that the number of anomaly samples in MVTecAD is considerably smaller than the number of normal samples, the discard rate is set to 0.3 rather than 0.5 as in other experiments. This setup reflects the filter's ability to discard noise samples. The results indicate that performance is limited when using a single layer for the patch feature descriptor, while multi-layer fusion enhances performance.
\\
\textbf{Additional experiments on hyperparameter robustness and boundary conditions.}
We present additional experiments on hyperparameter selection in Table~\ref{tab:ab_tau}, \ref{tab:ab_Tp}, and Table~\ref{tab:ab_num}, demonstrating strong robustness across all configurations. In the robustness experiments on the threshold $T_p$ in Table~\ref{tab:ab_Tp}, we fixed the upper limit at the 99th percentile to avoid excessively small sampling. Therefore, we considered only the maximum value percentage as a hyperparameter. Table \ref{tab:ab_num} details our exploration of methods for selecting the number of materials in a single synthetic image. The methods include random selection from 1 to 3, fixed selection of two, gradual increment with the number of iterations (from 1 to 10), and gradual increment with the number of iterations (from 1 to 5).
To explore the model's performance under boundary conditions, we reduced the number of normal samples in the training set to be equal to the number of abnormal samples based on the $\alpha=0.4$ setting, thereby creating a benchmark with a 50\% real noise rate. Table~\ref{noise50} presents the experimental results under this benchmark, indicating that our model still outperforms other methods.

\begin{table}[htbp]
    \centering
    \caption{Hyperparameter robustness experiment for the number of anomaly materials used in a single synthetic image, presenting I-AUROC/P-AUPRO(\%) results on the Real-IAD dataset.}
    \begin{tabular}{l|ccc}
        \toprule
        Synthetic number            & $\alpha = 0.0$ & $\alpha = 0.4$ \\
        \toprule
        rand(1,3)                   & 92.3/90.3      & 90.7/90.8      \\
        2                           & 92.4/90.4      & 90.8/91.0      \\
        from 1 to 10                & 92.3/90.4      & 90.6/91.3      \\
        \rowcolor{bluee}from 1 to 5 & 92.4/90.7      & 90.9/91.1      \\
        \bottomrule
    \end{tabular}
    \label{tab:ab_num}
\end{table}

\begin{table}[htbp]
    \centering
    \caption{Boundary experiment with 50\% real noise ratio, presenting I-AUROC/P-AUPRO(\%) results on the Real-IAD dataset.}
    \begin{tabular}{c|c|c}
        \toprule
        RD        & SoftPatch & SSFilter$_{448392}$ \\
        \toprule
        85.9/94.0 & 86.0/88.4 & 88.9/93.9           \\
        \bottomrule
    \end{tabular}
    \label{noise50}
\end{table}

\section{Conclusions}
\label{sec:conclu}
This paper introduces SSFilter, the first method to apply sample-level filtering in noisy unsupervised anomaly detection. SSFilter integrates two complementary filtering mechanisms for robust noise reduction and can identify potential noise signals, providing valuable real-world supervision for unsupervised models. In addition to achieving strong performance in end-to-end fully unsupervised anomaly detection, SSFilter enables dataset-wide filtering using a trained model, effectively bridging the gap between UAD and FUAD methods. 

\textbf{limitations.}
Despite SSFilter’s strong performance in noisy scenarios, its filtering mechanism discards some normal samples, causing a degree of performance degradation. Data-efficient unsupervised anomaly detection methods could improve our approach in future work.
{
    \small
    \bibliographystyle{ieeenat_fullname}
    \bibliography{main}
}
\clearpage
\appendix

\setcounter{table}{0}
\setcounter{figure}{0}
\renewcommand{\thefigure}{\Alph{section}.\arabic{figure}}
\renewcommand{\thetable}{\Alph{section}.\arabic{table}}
\setcounter{equation}{0}
\counterwithin{equation}{section}
\begin{onecolumn}
\section{More Quantitative Results for Each Category.}
Tables \ref{tab:mv-fuiad-0}, \ref{tab:mv-fuiad-1}, \ref{tab:mv-fuiad-2}, and \ref{tab:mv-fuiad-4} present the detailed results of SSFilter and various state-of-the-art (SoTA) methods for each category under four noise levels, offering comprehensive performance references.

\begin{table*}[htb]
    \centering
    \caption{FUAD performance~(I-AUROC/P-AUPRO\&I-AUROC$^\dagger$/P-AUPRO$^\dagger$) comparisons with state-of-the-art anomaly detection methods on Real-IAD with a noisy ratio of 0.0.}
    \label{tab:mv-fuiad-0}
    \renewcommand{\arraystretch}{1.2}
    \setlength\tabcolsep{6.0pt}
    \resizebox{1.0\linewidth}{!}{
        \begin{tabular}{ll | cc | c | cccc}
            \hline
            \multicolumn{2}{c|}{\multirow{2}{*}{Category}} & \multicolumn{2}{c|}{Embedding-based} & \multicolumn{1}{c|}{Synthetic-based} & \multicolumn{4}{c}{Reconstruction-based}                                                                                                                                    \\
                                                           &                                      & \textbf{PatchCore}                   & \textbf{SoftPatch}                       & \textbf{DeSTSeg}     & \textbf{RD}          & \textbf{Dinomaly}    & \textbf{SSFilter$_{256224}$} & \textbf{SSFilter$_{448392}$} \\
            \hline
                                                           & Audiojack                            & 90.1/88.1\&91.2/88.9                 & 90.6/88.3\&90.6/87.4                     & 91.4/94.9\&90.0/97.2 & 90.1/92.6\&90.5/92.7 & 93.4/97.2\&92.6/97.0 & 92.4/89.2\&-/-               & 92.3/94.7\&-/-               \\
                                                           & Bottle Cap                           & 96.8/96.5\&96.5/95.7                 & 97.8/98.2\&96.9/96.8                     & 94.5/99.8\&88.8/99.7 & 97.1/98.9\&96.7/98.6 & 96.3/99.0\&94.7/98.7 & 97.1/97.3\&-/-               & 95.4/98.6\&-/-               \\
                                                           & Button Battery                       & 86.7/83.7\&78.3/73.8                 & 88.4/83.6\&78.5/71.8                     & 88.6/92.0\&85.1/90.7 & 88.5/93.5\&77.7/85.7 & 89.0/93.2\&80.4/87.8 & 84.8/81.0\&-/-               & 82.5/87.3\&-/-               \\
                                                           & End Cap                              & 89.0/88.9\&89.2/90.5                 & 87.6/90.8\&86.5/89.5                     & 85.9/93.3\&85.0/93.1 & 87.0/93.1\&86.2/93.2 & 93.4/97.4\&92.6/97.3 & 89.3/91.6\&-/-               & 90.2/96.3\&-/-               \\
                                                           & Eraser                               & 95.1/95.0\&94.5/95.0                 & 94.6/96.5\&94.3/95.3                     & 91.6/96.7\&94.0/98.8 & 92.8/95.5\&92.9/94.8 & 95.5/98.6\&95.2/98.9 & 94.9/96.1\&-/-               & 95.0/98.9\&-/-               \\
                                                           & Fire Hood                            & 86.0/86.9\&86.9/86.9                 & 86.7/88.6\&85.3/88.2                     & 90.1/95.5\&91.4/96.5 & 87.1/93.1\&87.2/93.6 & 90.0/96.6\&90.1/96.9 & 88.5/90.7\&-/-               & 92.3/97.0\&-/-               \\
                                                           & Mint                                 & 79.2/70.1\&76.8/71.4                 & 76.7/72.1\&76.1/70.5                     & 83.7/87.8\&79.6/89.9 & 77.4/84.3\&75.6/83.2 & 88.0/81.2\&84.2/84.2 & 81.0/64.6\&-/-               & 83.3/81.7\&-/-               \\
                                                           & Mounts                               & 90.9/83.9\&90.5/85.8                 & 89.8/85.4\&89.7/84.9                     & 86.9/87.0\&85.5/90.6 & 91.0/91.0\&52.7/72.1 & 90.7/95.4\&90.7/95.0 & 88.9/90.0\&-/-               & 88.0/96.5\&-/-               \\
                                                           & PCB                                  & 94.3/90.8\&93.5/88.8                 & 93.5/91.3\&92.7/92.2                     & 96.0/95.5\&88.7/97.8 & 94.0/96.1\&94.3/96.1 & 96.8/97.9\&96.6/98.0 & 95.1/92.8\&-/-               & 95.5/97.5\&-/-               \\
                                                           & Phone Battery                        & 93.2/93.5\&93.8/93.8                 & 93.4/95.1\&92.4/94.8                     & 89.4/95.5\&90.6/96.0 & 93.6/98.5\&93.4/98.0 & 94.8/98.3\&94.7/98.5 & 96.1/96.8\&-/-               & 97.0/98.4\&-/-               \\
                                                           & Plastic Nut                          & 95.5/95.9\&92.8/96.3                 & 93.5/96.2\&91.4/95.3                     & 87.4/98.6\&84.7/98.0 & 93.8/97.9\&91.0/98.0 & 96.4/98.4\&94.1/98.2 & 94.0/94.1\&-/-               & 93.3/98.0\&-/-               \\
                                                           & Plastic Plug                         & 92.5/91.2\&92.5/91.9                 & 92.0/93.1\&91.7/92.7                     & 90.2/91.8\&83.3/85.6 & 94.3/96.0\&94.0/96.1 & 94.4/95.4\&93.8/95.5 & 93.0/90.0\&-/-               & 94.1/97.0\&-/-               \\
                                                           & Porcelain Doll                       & 90.3/90.2\&90.2/90.5                 & 89.1/91.7\&88.7/91.8                     & 90.7/97.7\&90.2/97.3 & 91.6/95.8\&90.9/96.2 & 91.3/97.1\&91.3/97.4 & 92.6/92.2\&-/-               & 93.9/98.2\&-/-               \\
                                                           & Regulator                            & 87.3/91.2\&75.5/83.5                 & 84.7/93.1\&68.7/79.5                     & 91.9/95.5\&88.2/91.7 & 89.7/97.5\&77.8/85.4 & 92.1/97.8\&78.6/92.9 & 80.5/79.1\&-/-               & 81.4/91.0\&-/-               \\
                                                           & Rolled Strip Base                    & 99.4/98.5\&99.3/98.7                 & 99.3/98.7\&99.2/98.7                     & 97.7/99.4\&98.4/99.7 & 99.7/99.0\&99.7/99.2 & 99.4/98.6\&99.4/98.7 & 99.6/97.9\&-/-               & 99.2/98.6\&-/-               \\
                                                           & SIM Card Set                         & 98.4/89.5\&98.3/90.0                 & 98.2/90.8\&98.3/90.7                     & 96.2/98.1\&95.7/97.3 & 97.5/90.9\&96.9/91.4 & 98.2/94.5\&98.3/95.8 & 98.9/92.6\&-/-               & 98.5/96.4\&-/-               \\
                                                           & Switch                               & 96.3/93.1\&96.2/92.5                 & 95.5/94.7\&95.7/93.8                     & 98.1/97.6\&97.3/98.6 & 96.2/96.9\&95.9/96.5 & 99.0/97.9\&98.9/97.6 & 96.5/95.0\&-/-               & 97.5/95.8\&-/-               \\
                                                           & Tape                                 & 98.8/98.2\&98.8/98.3                 & 98.7/98.7\&98.6/98.5                     & 97.8/99.5\&97.6/99.5 & 98.6/99.0\&98.7/99.0 & 99.1/99.2\&99.0/99.3 & 98.0/98.2\&-/-               & 98.4/98.9\&-/-               \\
                                                           & Terminal Block                       & 97.5/96.7\&96.7/96.2                 & 96.2/96.8\&96.2/96.9                     & 92.5/98.9\&93.6/98.5 & 98.2/98.5\&97.9/98.5 & 98.2/98.6\&98.1/98.8 & 97.6/97.8\&-/-               & 97.9/99.2\&-/-               \\
                                                           & Toothbrush                           & 89.9/89.6\&87.6/88.4                 & 89.2/91.3\&87.2/89.3                     & 87.9/81.3\&89.4/89.9 & 85.1/90.4\&79.9/87.7 & 88.6/87.2\&84.2/83.9 & 86.1/85.9\&-/-               & 82.0/84.0\&-/-               \\
                                                           & Toy                                  & 89.7/87.2\&89.1/87.7                 & 88.4/89.2\&88.1/88.0                     & 84.1/93.3\&83.7/92.3 & 89.0/91.4\&89.0/91.9 & 93.0/93.4\&92.0/93.3 & 91.7/86.8\&-/-               & 86.4/84.9\&-/-               \\
                                                           & Toy Brick                            & 83.8/82.4\&82.5/81.6                 & 82.9/82.4\&81.0/81.5                     & 78.5/83.0\&81.1/83.5 & 75.6/82.8\&74.6/85.0 & 80.2/84.1\&79.9/84.0 & 83.8/80.4\&-/-               & 91.5/92.9\&-/-               \\
                                                           & Transistor1                          & 98.1/96.9\&98.1/96.8                 & 97.7/96.9\&97.4/97.1                     & 97.3/95.7\&96.7/94.2 & 98.5/98.4\&98.4/98.3 & 98.9/98.3\&98.6/98.1 & 97.9/95.8\&-/-               & 97.0/97.4\&-/-               \\
                                                           & USB                                  & 94.9/96.7\&95.0/95.9                 & 93.6/96.6\&93.7/95.7                     & 94.8/98.3\&94.7/99.0 & 95.3/98.4\&95.2/98.5 & 96.6/98.7\&96.3/98.8 & 94.8/97.6\&-/-               & 92.7/91.6\&-/-               \\
                                                           & USB Adaptor                          & 88.0/82.7\&87.5/82.1                 & 88.1/84.3\&87.7/84.3                     & 80.5/96.7\&77.6/95.2 & 88.6/82.5\&88.2/83.3 & 90.0/92.0\&90.4/92.8 & 91.6/81.0\&-/-               & 96.0/98.1\&-/-               \\
                                                           & U Block                              & 94.1/95.0\&94.0/95.0                 & 93.4/94.5\&93.1/95.0                     & 93.4/99.2\&92.7/98.7 & 94.7/98.2\&93.8/97.9 & 95.4/98.0\&95.0/97.8 & 93.5/95.2\&-/-               & 92.7/97.8\&-/-               \\
                                                           & Vcpill                               & 93.2/89.6\&93.2/89.2                 & 92.4/89.9\&91.3/89.6                     & 93.4/95.5\&94.1/94.0 & 91.7/91.1\&91.5/91.7 & 95.5/96.3\&95.2/96.3 & 93.8/92.6\&-/-               & 93.7/95.1\&-/-               \\
                                                           & Wooden Beads                         & 89.6/83.5\&89.1/83.4                 & 90.5/86.3\&89.8/87.0                     & 91.3/90.3\&90.1/92.5 & 88.7/90.9\&88.9/90.8 & 91.2/93.2\&90.6/94.0 & 91.9/89.5\&-/-               & 91.1/94.2\&-/-               \\
                                                           & Woodstick                            & 80.7/71.7\&80.1/70.7                 & 80.3/71.5\&80.7/71.7                     & 92.8/91.3\&89.6/91.6 & 88.1/92.3\&87.4/92.1 & 92.6/94.7\&92.3/93.9 & 91.0/90.9\&-/-               & 92.2/94.7\&-/-               \\
                                                           & Zipper                               & 98.1/96.1\&98.0/95.7                 & 97.6/95.7\&96.8/94.8                     & 99.1/96.0\&99.1/96.1 & 99.7/98.4\&99.3/98.2 & 99.0/97.8\&98.5/97.9 & 96.3/97.0\&-/-               & 95.3/97.6\&-/-               \\
            \hline
                                                           & Average All                          & 91.9/89.8\&90.9/89.2                 & 91.4/90.7\&89.9/89.4                     & 91.1/94.5\&89.9/94.8 & 91.8/94.1\&89.2/92.8 & 93.9/95.5\&92.5/95.2 & 92.4/90.7\&-/-               & 92.5/94.9\&-/-               \\
            \hline
        \end{tabular}
    }
\end{table*}

\begin{table*}[htb]
    \centering
    \caption{FUAD performance~(I-AUROC/P-AUPRO\&I-AUROC$^\dagger$/P-AUPRO$^\dagger$) comparisons with state-of-the-art anomaly detection methods on Real-IAD with a noisy ratio of 0.1.}
    \label{tab:mv-fuiad-1}
    \renewcommand{\arraystretch}{1.2}
    \setlength\tabcolsep{6.0pt}
    \resizebox{1.0\linewidth}{!}{
        \begin{tabular}{ll | cc | c | cccc}
            \hline
            \multicolumn{2}{c|}{\multirow{2}{*}{Category}} & \multicolumn{2}{c|}{Embedding-based} & \multicolumn{1}{c|}{Synthetic-based} & \multicolumn{4}{c}{Reconstruction-based}                                                                                                                                    \\
                                                           &                                      & \textbf{PatchCore}                   & \textbf{SoftPatch}                       & \textbf{DeSTSeg}     & \textbf{RD}          & \textbf{Dinomaly}    & \textbf{SSFilter$_{256224}$} & \textbf{SSFilter$_{448392}$} \\
            \hline
                                                           & Audiojack                            & 89.4/85.9\&90.3/88.0                 & 88.9/89.4\&89.0/89.5                     & 93.2/96.3\&93.4/97.6 & 89.8/92.7\&89.5/93.1 & 90.2/96.9\&92.3/97.2 & 92.5/90.7\&-/-               & 91.9/94.9\&-/-               \\
                                                           & Bottle Cap                           & 95.3/95.7\&96.5/96.7                 & 97.6/98.4\&97.2/97.2                     & 95.7/99.9\&94.8/99.9 & 96.0/99.0\&96.9/98.8 & 91.5/98.9\&94.5/98.8 & 97.0/97.3\&-/-               & 96.1/98.6\&-/-               \\
                                                           & Button Battery                       & 80.9/85.0\&76.3/79.1                 & 80.6/85.2\&73.7/73.8                     & 86.9/87.2\&85.3/92.0 & 85.6/92.9\&77.3/87.3 & 83.7/93.6\&79.6/88.0 & 82.4/84.6\&-/-               & 82.3/89.1\&-/-               \\
                                                           & End Cap                              & 84.1/83.3\&87.5/86.1                 & 87.2/89.1\&87.1/90.7                     & 82.9/90.6\&82.2/89.6 & 82.3/92.8\&84.5/92.9 & 88.6/96.9\&91.1/97.0 & 88.5/91.6\&-/-               & 90.1/95.7\&-/-               \\
                                                           & Eraser                               & 93.1/95.3\&94.3/95.9                 & 94.3/95.6\&94.1/96.0                     & 89.9/98.0\&93.0/98.7 & 91.5/96.4\&92.3/96.1 & 94.0/98.7\&95.5/99.0 & 95.6/96.0\&-/-               & 96.2/99.0\&-/-               \\
                                                           & Fire Hood                            & 83.9/85.8\&86.6/87.0                 & 85.9/88.6\&85.7/88.5                     & 91.6/96.5\&93.0/96.1 & 85.2/93.5\&87.2/93.8 & 88.5/96.8\&90.4/96.0 & 88.9/91.0\&-/-               & 93.1/97.1\&-/-               \\
                                                           & Mint                                 & 77.1/71.7\&75.5/70.4                 & 75.9/71.9\&75.2/72.6                     & 81.8/84.5\&81.7/86.9 & 75.6/85.0\&75.8/85.3 & 83.7/86.4\&82.3/85.1 & 79.9/71.5\&-/-               & 83.0/81.0\&-/-               \\
                                                           & Mounts                               & 93.0/83.7\&90.6/86.3                 & 90.7/85.0\&89.8/84.3                     & 85.9/93.7\&85.9/93.4 & 92.3/92.5\&90.8/92.7 & 92.3/95.8\&90.6/95.1 & 88.1/89.1\&-/-               & 87.1/95.7\&-/-               \\
                                                           & PCB                                  & 92.3/91.1\&94.0/91.0                 & 93.1/91.3\&92.8/91.1                     & 93.8/97.9\&85.5/97.3 & 93.4/96.2\&93.7/96.0 & 95.6/97.9\&96.2/97.8 & 94.8/92.4\&-/-               & 95.5/96.8\&-/-               \\
                                                           & Phone Battery                        & 93.3/94.8\&93.7/94.5                 & 93.0/95.6\&92.5/95.7                     & 93.1/96.9\&90.0/96.8 & 93.3/98.9\&93.8/98.6 & 94.9/98.8\&95.9/98.8 & 96.7/97.1\&-/-               & 96.9/98.9\&-/-               \\
                                                           & Plastic Nut                          & 94.2/94.8\&93.0/95.1                 & 93.2/96.1\&91.9/95.8                     & 90.7/98.7\&86.8/98.9 & 92.8/98.0\&90.9/98.0 & 93.8/98.6\&94.5/98.5 & 94.5/96.1\&-/-               & 94.5/98.0\&-/-               \\
                                                           & Plastic Plug                         & 91.9/92.6\&92.3/92.5                 & 92.4/93.3\&91.6/92.6                     & 88.8/96.1\&89.6/95.9 & 93.4/96.9\&93.6/96.3 & 93.6/97.3\&94.2/96.7 & 93.4/91.4\&-/-               & 94.2/97.2\&-/-               \\
                                                           & Porcelain Doll                       & 89.4/90.8\&90.3/91.1                 & 88.1/92.7\&88.9/92.7                     & 80.6/95.1\&89.6/97.0 & 90.7/96.6\&91.0/96.5 & 91.5/98.3\&93.0/98.3 & 92.8/91.8\&-/-               & 94.4/98.1\&-/-               \\
                                                           & Regulator                            & 85.3/91.5\&74.2/84.0                 & 82.7/92.5\&69.1/79.3                     & 88.4/90.7\&86.9/96.2 & 89.6/98.3\&77.6/86.4 & 86.9/98.8\&77.6/93.5 & 81.4/81.4\&-/-               & 80.8/90.6\&-/-               \\
                                                           & Rolled Strip Base                    & 98.7/98.3\&99.3/98.8                 & 99.3/98.8\&99.1/98.9                     & 97.3/99.3\&98.3/99.7 & 99.3/99.5\&99.7/99.3 & 97.7/99.0\&99.4/98.7 & 99.5/98.5\&-/-               & 99.4/99.2\&-/-               \\
                                                           & SIM Card Set                         & 96.7/95.2\&97.7/94.3                 & 96.8/95.6\&97.9/93.6                     & 83.9/94.7\&92.5/98.0 & 95.8/96.4\&96.3/94.3 & 96.6/97.6\&98.0/97.5 & 98.4/94.7\&-/-               & 98.7/97.9\&-/-               \\
                                                           & Switch                               & 94.8/89.1\&95.4/92.7                 & 95.3/93.5\&95.5/94.0                     & 96.5/93.8\&97.3/98.3 & 96.1/96.1\&96.0/96.7 & 98.5/96.8\&98.7/97.4 & 96.4/95.1\&-/-               & 97.7/95.8\&-/-               \\
                                                           & Tape                                 & 97.9/97.6\&98.9/98.2                 & 98.7/98.8\&98.7/98.6                     & 97.2/99.4\&98.1/99.6 & 98.2/98.8\&98.4/98.9 & 97.9/99.3\&98.8/99.2 & 98.2/98.0\&-/-               & 98.4/99.1\&-/-               \\
                                                           & Terminal Block                       & 96.4/95.8\&97.4/96.5                 & 96.6/97.3\&96.1/96.7                     & 91.2/97.4\&92.1/99.0 & 98.3/98.9\&98.1/98.8 & 97.6/99.2\&98.4/98.8 & 97.6/97.7\&-/-               & 97.9/99.0\&-/-               \\
                                                           & Toothbrush                           & 89.6/89.0\&88.7/89.3                 & 90.1/91.0\&88.5/90.4                     & 86.4/76.4\&89.7/89.6 & 81.8/90.8\&80.2/89.1 & 87.6/88.7\&86.2/86.9 & 87.8/86.4\&-/-               & 83.7/85.2\&-/-               \\
                                                           & Toy                                  & 85.7/86.2\&88.0/87.0                 & 86.9/89.0\&88.6/86.9                     & 83.1/89.2\&82.7/89.2 & 85.7/91.7\&88.6/92.1 & 88.7/93.0\&92.2/93.5 & 90.4/89.0\&-/-               & 83.6/84.8\&-/-               \\
                                                           & Toy Brick                            & 80.4/80.6\&80.8/80.9                 & 82.3/83.5\&80.7/82.4                     & 71.9/80.6\&74.3/79.0 & 71.1/81.2\&74.9/84.4 & 72.3/80.6\&76.6/80.4 & 84.7/81.1\&-/-               & 92.0/93.0\&-/-               \\
                                                           & Transistor1                          & 95.2/94.1\&97.7/96.3                 & 95.9/96.1\&97.5/96.9                     & 90.1/88.3\&95.2/94.9 & 96.1/98.2\&97.8/98.3 & 97.3/98.1\&98.1/98.2 & 97.1/96.3\&-/-               & 97.3/97.4\&-/-               \\
                                                           & USB                                  & 93.2/93.5\&95.2/96.8                 & 94.1/95.9\&93.6/96.4                     & 93.6/97.8\&95.2/96.9 & 95.3/98.5\&95.1/98.4 & 94.7/98.3\&96.2/98.7 & 95.2/97.6\&-/-               & 93.2/95.5\&-/-               \\
                                                           & USB Adaptor                          & 84.1/82.0\&87.1/83.8                 & 87.2/84.5\&87.6/84.2                     & 87.2/97.2\&87.5/96.2 & 82.8/91.4\&87.2/88.7 & 86.7/96.6\&91.2/96.6 & 90.8/86.7\&-/-               & 95.9/98.2\&-/-               \\
                                                           & U Block                              & 94.0/94.4\&94.5/95.1                 & 94.0/94.5\&92.7/94.9                     & 90.1/97.7\&94.0/98.7 & 93.9/98.6\&93.5/98.4 & 94.4/98.3\&95.4/98.3 & 93.6/95.6\&-/-               & 93.2/97.9\&-/-               \\
                                                           & Vcpill                               & 92.4/89.1\&93.0/90.5                 & 91.5/89.9\&91.5/89.3                     & 91.2/95.5\&93.3/92.3 & 89.2/91.1\&90.6/91.2 & 93.1/95.9\&94.9/96.2 & 93.5/92.7\&-/-               & 93.7/94.9\&-/-               \\
                                                           & Wooden Beads                         & 86.3/80.5\&88.8/83.3                 & 89.4/87.1\&90.1/85.7                     & 77.5/88.9\&89.6/91.1 & 86.5/90.6\&87.4/91.1 & 88.3/94.0\&90.2/94.3 & 91.7/89.2\&-/-               & 91.6/94.4\&-/-               \\
                                                           & Woodstick                            & 78.9/67.2\&80.3/68.5                 & 80.5/73.2\&79.8/72.5                     & 92.5/95.3\&91.2/93.8 & 86.0/91.4\&87.9/91.7 & 89.5/94.8\&91.5/93.8 & 91.0/91.0\&-/-               & 92.3/94.5\&-/-               \\
                                                           & Zipper                               & 97.3/95.0\&98.2/95.7                 & 97.2/95.5\&97.5/95.2                     & 94.7/85.3\&99.2/96.2 & 99.3/98.4\&99.6/98.6 & 98.2/97.7\&98.8/97.8 & 96.3/96.4\&-/-               & 95.3/97.7\&-/-               \\
            \hline
                                                           & Average All                          & 90.2/89.0\&90.5/89.5                 & 90.7/91.0\&89.8/89.9                     & 88.9/93.3\&90.3/95.0 & 90.2/94.7\&90.3/94.1 & 91.6/96.1\&92.4/95.5 & 92.3/91.6\&-/-               & 92.7/95.2\&-/-               \\
            \hline
        \end{tabular}
    }
\end{table*}

\begin{table*}[htb]
    \centering
    \caption{FUAD performance~(I-AUROC/P-AUPRO\&I-AUROC$^\dagger$/P-AUPRO$^\dagger$) comparisons with state-of-the-art anomaly detection methods on Real-IAD with a noisy ratio of 0.2.}
    \label{tab:mv-fuiad-2}
    \renewcommand{\arraystretch}{1.2}
    \setlength\tabcolsep{6.0pt}
    \resizebox{1.0\linewidth}{!}{
        \begin{tabular}{ll | cc | c | cccc}
            \hline
            \multicolumn{2}{c|}{\multirow{2}{*}{Category}} & \multicolumn{2}{c|}{Embedding-based} & \multicolumn{1}{c|}{Synthetic-based} & \multicolumn{4}{c}{Reconstruction-based}                                                                                                                                    \\
                                                           &                                      & \textbf{PatchCore}                   & \textbf{SoftPatch}                       & \textbf{DeSTSeg}     & \textbf{RD}          & \textbf{Dinomaly}    & \textbf{SSFilter$_{256224}$} & \textbf{SSFilter$_{448392}$} \\
            \hline
                                                           & Audiojack                            & 87.6/87.7\&88.3/87.9                 & 87.7/88.5\&88.6/89.0                     & 90.1/94.4\&93.3/97.8 & 88.6/93/2\&89.8/92.8 & 88.6/96.7\&90.9/96.7 & 92.3/89.9\&-/-               & 91.4/94.7\&-/-               \\
                                                           & Bottle Cap                           & 92.9/95.2\&95.8/95.3                 & 97.4/98.3\&97.0/98.1                     & 92.1/99.2\&95.5/99.9 & 95.2/98/9\&96.4/98.6 & 90.2/98.8\&93.4/99.0 & 96.9/97.1\&-/-               & 95.7/98.5\&-/-               \\
                                                           & Button Battery                       & 80.4/84.1\&79.4/84.2                 & 80.6/86.4\&77.1/84.2                     & 86.2/92.9\&86.9/90.1 & 84.9/92/8\&80.9/90.8 & 82.0/93.3\&80.5/89.9 & 81.4/82.9\&-/-               & 81.7/89.8\&-/-               \\
                                                           & End Cap                              & 82.7/84.1\&85.1/85.4                 & 86.6/90.3\&86.0/88.9                     & 79.2/89.7\&76.0/90.5 & 80.5/92/0\&83.0/92.6 & 86.1/96.6\&89.6/97.0 & 86.8/90.0\&-/-               & 88.8/96.4\&-/-               \\
                                                           & Eraser                               & 92.1/95.3\&94.0/95.2                 & 94.2/96.6\&94.4/95.8                     & 86.6/96.7\&92.8/98.1 & 91.7/97/1\&92.2/96.8 & 92.6/98.6\&95.4/98.9 & 95.1/95.6\&-/-               & 95.9/98.9\&-/-               \\
                                                           & Fire Hood                            & 82.5/85.3\&85.8/87.7                 & 85.1/88.2\&86.0/87.9                     & 79.0/89.3\&87.1/96.1 & 84.1/93/5\&86.3/93.4 & 87.2/96.4\&90.4/96.4 & 89.4/91.3\&-/-               & 93.6/97.3\&-/-               \\
                                                           & Mint                                 & 76.9/70.1\&74.6/69.5                 & 75.8/73.3\&73.6/70.9                     & 82.7/87.5\&83.9/94.1 & 74.3/86/1\&74.7/84.6 & 81.8/82.0\&80.3/83.4 & 79.8/72.7\&-/-               & 82.5/81.8\&-/-               \\
                                                           & Mounts                               & 92.3/83.8\&90.4/84.4                 & 92.4/85.9\&89.7/84.8                     & 86.5/94.9\&87.5/96.8 & 92.1/93/8\&90.2/92.0 & 92.1/95.4\&90.5/95.2 & 88.3/89.3\&-/-               & 87.1/95.8\&-/-               \\
                                                           & PCB                                  & 92.1/92.0\&93.0/92.2                 & 91.9/90.9\&92.3/91.7                     & 90.8/93.2\&94.3/98.2 & 92.9/96/2\&93.1/95.7 & 94.5/97.8\&95.7/98.0 & 94.4/91.8\&-/-               & 95.4/96.9\&-/-               \\
                                                           & Phone Battery                        & 90.9/94.1\&93.6/95.0                 & 92.8/95.1\&92.0/95.3                     & 91.8/96.7\&91.8/97.1 & 92.2/98/5\&93.4/98.8 & 93.0/98.8\&95.9/98.8 & 96.7/98.1\&-/-               & 97.1/98.8\&-/-               \\
                                                           & Plastic Nut                          & 93.1/94.7\&93.8/95.6                 & 93.1/96.2\&91.7/95.1                     & 85.1/96.6\&87.5/96.6 & 91.7/98/2\&92.5/97.9 & 92.3/98.2\&94.4/98.7 & 94.4/96.0\&-/-               & 94.8/97.7\&-/-               \\
                                                           & Plastic Plug                         & 90.6/91.9\&92.1/92.2                 & 92.3/93.3\&91.5/91.7                     & 84.4/94.0\&91.0/96.8 & 91.8/96/8\&93.1/96.9 & 92.3/97.1\&93.9/97.4 & 93.5/91.1\&-/-               & 94.1/97.3\&-/-               \\
                                                           & Porcelain Doll                       & 87.7/91.2\&89.2/91.9                 & 87.7/92.1\&88.7/92.9                     & 86.9/96.1\&89.7/97.9 & 90.3/96/9\&91.2/96.3 & 90.6/98.2\&92.8/98.5 & 92.5/92.0\&-/-               & 93.8/98.2\&-/-               \\
                                                           & Regulator                            & 82.9/90.4\&77.1/88.0                 & 83.4/94.4\&71.0/83.9                     & 90.5/95.2\&87.8/85.8 & 88.4/98/4\&82.2/93.3 & 85.0/98.5\&79.6/96.0 & 81.6/87.6\&-/-               & 82.6/92.2\&-/-               \\
                                                           & Rolled Strip Base                    & 96.7/97.5\&99.2/98.6                 & 98.4/98.3\&99.2/98.9                     & 95.0/99.1\&97.8/99.1 & 98.5/99/5\&99.7/99.3 & 95.6/98.8\&99.1/98.7 & 99.5/98.6\&-/-               & 99.4/99.3\&-/-               \\
                                                           & SIM Card Set                         & 95.8/94.9\&97.5/94.7                 & 96.2/95.2\&97.6/93.8                     & 90.8/98.1\&95.6/99.0 & 95.3/96/7\&96.1/94.8 & 95.7/97.8\&97.9/97.6 & 98.3/94.3\&-/-               & 98.8/97.9\&-/-               \\
                                                           & Switch                               & 94.0/89.2\&95.1/93.3                 & 95.7/93.8\&95.6/93.7                     & 90.9/91.8\&96.9/96.5 & 95.8/95/6\&95.7/96.5 & 98.0/96.4\&98.2/97.3 & 96.1/94.7\&-/-               & 97.5/95.5\&-/-               \\
                                                           & Tape                                 & 97.2/98.0\&98.9/98.5                 & 98.4/98.5\&98.7/98.6                     & 94.3/98.5\&98.6/99.7 & 97.7/99/0\&98.6/98.9 & 96.9/99.0\&98.9/99.2 & 98.2/98.0\&-/-               & 98.4/99.0\&-/-               \\
                                                           & Terminal Block                       & 95.3/94.5\&97.2/96.1                 & 96.8/96.9\&96.3/96.9                     & 90.2/98.0\&96.0/99.5 & 98.0/99/1\&98.5/99.0 & 96.7/99.1\&98.3/98.8 & 97.5/97.6\&-/-               & 98.3/99.0\&-/-               \\
                                                           & Toothbrush                           & 87.4/88.3\&87.4/86.9                 & 89.3/89.9\&88.4/89.9                     & 78.8/67.2\&86.5/82.1 & 80.7/90/7\&80.0/89.1 & 84.5/87.8\&83.8/85.5 & 85.8/86.8\&-/-               & 83.0/84.7\&-/-               \\
                                                           & Toy                                  & 82.6/85.3\&85.4/85.0                 & 85.9/87.8\&86.1/88.6                     & 70.0/58.6\&81.2/88.8 & 84.6/92/2\&87.2/92.8 & 85.2/92.4\&90.8/93.5 & 89.8/87.3\&-/-               & 84.0/82.8\&-/-               \\
                                                           & Toy Brick                            & 78.3/78.1\&80.9/79.3                 & 80.2/82.8\&81.3/83.6                     & 69.0/77.8\&76.6/78.6 & 71.2/81/2\&73.0/82.5 & 68.5/78.4\&75.0/82.3 & 85.0/79.4\&-/-               & 91.2/91.0\&-/-               \\
                                                           & Transistor1                          & 93.4/91.3\&97.0/95.2                 & 94.3/95.1\&97.0/96.8                     & 87.0/90.4\&96.1/94.5 & 94.7/97/8\&97.2/98.1 & 95.6/97.7\&97.4/97.8 & 96.7/96.2\&-/-               & 97.1/97.2\&-/-               \\
                                                           & USB                                  & 92.0/92.5\&94.4/95.3                 & 93.2/95.8\&93.9/96.5                     & 90.8/96.6\&95.7/98.7 & 94.5/98/3\&95.0/98.5 & 94.1/98.3\&96.5/98.6 & 95.2/97.1\&-/-               & 93.3/95.1\&-/-               \\
                                                           & USB Adaptor                          & 82.6/81.3\&85.8/82.4                 & 85.2/85.5\&86.8/83.8                     & 86.2/96.8\&87.3/96.9 & 80.8/91/1\&85.3/90.8 & 83.3/96.9\&89.3/96.6 & 90.7/86.9\&-/-               & 96.4/98.1\&-/-               \\
                                                           & U Block                              & 93.3/94.6\&94.1/94.5                 & 94.2/95.9\&93.6/95.3                     & 82.2/96.1\&94.0/99.0 & 93.4/98/5\&93.4/98.2 & 92.7/98.2\&94.8/98.5 & 93.6/95.8\&-/-               & 92.9/97.6\&-/-               \\
                                                           & Vcpill                               & 90.6/87.7\&92.0/89.7                 & 90.3/89.0\&91.1/89.4                     & 85.9/92.3\&94.5/92.8 & 88.2/90/7\&91.0/91.4 & 92.4/95.2\&94.0/95.7 & 93.3/92.0\&-/-               & 93.4/95.2\&-/-               \\
                                                           & Wooden Beads                         & 85.7/80.6\&87.9/82.3                 & 87.3/85.4\&89.2/86.7                     & 82.8/87.2\&88.5/91.9 & 85.4/90/6\&87.4/91.0 & 86.3/93.1\&89.4/93.2 & 91.4/88.6\&-/-               & 91.0/94.2\&-/-               \\
                                                           & Woodstick                            & 75.5/63.0\&79.3/69.6                 & 81.1/69.9\&79.4/72.0                     & 87.9/89.9\&91.0/91.3 & 85.6/90/8\&87.6/91.6 & 87.3/91.4\&90.9/95.4 & 91.0/90.1\&-/-               & 92.4/93.8\&-/-               \\
                                                           & Zipper                               & 96.4/93.9\&98.0/95.9                 & 96.5/95.4\&97.5/95.6                     & 94.7/90.3\&98.8/96.0 & 99.2/98/3\&99.6/98.6 & 97.9/97.4\&98.9/98.2 & 96.8/96.9\&-/-               & 96.2/98.1\&-/-               \\
            \hline
                                                           & Average All                          & 88.7/88.3\&90.1/89.4                 & 90.1/90.8\&89.7/90.3                     & 86.3/91.5\&90.7/94.7 & 89.4/94.8\&90.1/94.4 & 90.0/95.5\&91.9/95.7 & 92.1/91.5\&-/-               & 92.6/95.1\&-/-               \\
            \hline
        \end{tabular}
    }
\end{table*}

\begin{table*}[htb]
    \centering
    \caption{FUAD performance~(I-AUROC/P-AUPRO\&I-AUROC$^\dagger$/P-AUPRO$^\dagger$) comparisons with state-of-the-art anomaly detection methods on Real-IAD with a noisy ratio of 0.4.}
    \label{tab:mv-fuiad-4}
    \renewcommand{\arraystretch}{1.2}
    \setlength\tabcolsep{6.0pt}
    \resizebox{1.0\linewidth}{!}{
        \begin{tabular}{ll | cc | c | cccc}
            \hline
            \multicolumn{2}{c|}{\multirow{2}{*}{Category}} & \multicolumn{2}{c|}{Embedding-based} & \multicolumn{1}{c|}{Synthetic-based} & \multicolumn{4}{c}{Reconstruction-based}                                                                                                                                    \\
                                                           &                                      & \textbf{PatchCore}                   & \textbf{SoftPatch}                       & \textbf{DeSTSeg}     & \textbf{RD}          & \textbf{Dinomaly}    & \textbf{SSFilter$_{256224}$} & \textbf{SSFilter$_{448392}$} \\
            \hline
                                                           & Audiojack                            & 85.3/81.3\&86.6/86.7                 & 85.6/88.3\&86.4/88.6                     & 83.2/95.4\&90.3/95.4 & 85.9/92.5\&87.7/92.7 & 83.7/96.0\&87.7/96.1 & 90.6/89.5\&-/-               & 88.4/94.2\&-/-               \\
                                                           & Bottle Cap                           & 91.6/94.9\&95.6/96.3                 & 96.8/97.8\&97.6/98.2                     & 85.9/99.0\&95.2/99.6 & 92.7/98.5\&96.7/98.7 & 85.6/97.9\&92.0/98.5 & 96.3/96.9\&-/-               & 95.3/98.2\&-/-               \\
                                                           & Button Battery                       & 77.7/83.3\&77.4/84.3                 & 77.1/84.9\&77.4/84.7                     & 74.7/88.5\&85.9/91.0 & 82.6/92.5\&82.2/92.2 & 78.1/91.4\&79.1/92.2 & 79.7/86.2\&-/-               & 78.8/89.6\&-/-               \\
                                                           & End Cap                              & 79.2/80.2\&82.8/83.3                 & 82.9/86.2\&85.3/88.8                     & 77.4/83.2\&75.4/90.4 & 77.5/91.4\&81.8/91.8 & 82.9/96.3\&87.2/96.4 & 84.9/88.1\&-/-               & 88.0/95.9\&-/-               \\
                                                           & Eraser                               & 90.5/93.0\&93.6/94.1                 & 93.4/96.4\&94.0/96.0                     & 74.4/80.3\&93.7/99.0 & 88.6/97.1\&91.5/95.5 & 89.4/97.8\&94.5/98.9 & 95.4/94.9\&-/-               & 96.3/99.0\&-/-               \\
                                                           & Fire Hood                            & 80.3/82.4\&84.9/85.7                 & 84.0/87.9\&85.4/87.8                     & 89.1/96.2\&92.1/96.5 & 83.7/92.4\&85.8/92.8 & 85.8/95.5\&89.7/96.0 & 87.2/89.7\&-/-               & 93.1/97.1\&-/-               \\
                                                           & Mint                                 & 72.5/69.8\&75.8/69.7                 & 74.7/73.2\&75.1/70.1                     & 74.2/84.5\&79.9/82.7 & 70.7/83.6\&75.5/86.1 & 78.6/84.7\&79.8/84.3 & 78.1/72.1\&-/-               & 81.7/81.0\&-/-               \\
                                                           & Mounts                               & 90.7/82.4\&91.7/82.4                 & 90.9/84.2\&88.9/84.7                     & 77.1/88.5\&87.1/97.6 & 91.1/94.4\&90.1/92.1 & 90.7/95.4\&91.1/95.2 & 87.4/88.8\&-/-               & 86.9/95.9\&-/-               \\
                                                           & PCB                                  & 90.5/88.2\&91.2/90.7                 & 90.0/90.1\&91.0/90.3                     & 78.3/95.6\&90.7/98.5 & 91.9/95.6\&92.6/95.4 & 92.8/97.9\&94.1/97.8 & 93.1/92.6\&-/-               & 94.5/97.1\&-/-               \\
                                                           & Phone Battery                        & 88.1/90.7\&92.5/94.4                 & 91.0/95.3\&91.7/96.0                     & 87.5/91.3\&93.0/96.7 & 90.6/98.6\&93.1/98.6 & 91.2/98.7\&94.9/98.9 & 96.8/98.0\&-/-               & 97.0/98.5\&-/-               \\
                                                           & Plastic Nut                          & 92.4/93.3\&93.6/95.1                 & 91.4/95.8\&91.1/95.4                     & 85.6/98.2\&87.7/98.8 & 90.2/98.1\&92.2/98.1 & 90.3/97.9\&93.9/98.8 & 94.1/96.1\&-/-               & 95.1/97.7\&-/-               \\
                                                           & Plastic Plug                         & 89.7/91.1\&92.3/93.0                 & 91.8/93.4\&91.1/92.6                     & 77.5/92.8\&90.4/92.9 & 91.6/97.2\&93.2/96.6 & 90.3/96.7\&93.5/97.1 & 93.0/93.6\&-/-               & 93.7/97.4\&-/-               \\
                                                           & Porcelain Doll                       & 85.2/89.5\&88.8/91.2                 & 86.3/91.1\&87.8/91.9                     & 81.2/96.8\&88.1/94.9 & 89.5/96.9\&90.7/96.8 & 88.3/97.8\&92.1/98.3 & 92.0/91.6\&-/-               & 93.6/97.9\&-/-               \\
                                                           & Regulator                            & 81.2/90.6\&78.4/88.8                 & 82.8/91.4\&76.2/89.8                     & 87.2/96.6\&83.5/85.0 & 88.4/98.2\&85.1/97.1 & 82.4/98.3\&80.5/97.8 & 81.5/88.4\&-/-               & 81.9/93.9\&-/-               \\
                                                           & Rolled Strip Base                    & 94.0/96.4\&96.0/96.9                 & 95.1/97.1\&97.6/97.8                     & 92.2/97.4\&94.4/91.9 & 97.4/99.5\&97.9/99.2 & 92.4/98.5\&95.7/98.5 & 97.5/98.0\&-/-               & 98.0/98.9\&-/-               \\
                                                           & SIM Card Set                         & 94.0/94.4\&96.4/95.0                 & 95.3/95.6\&96.6/95.8                     & 88.7/98.1\&96.3/98.7 & 93.2/96.6\&95.2/95.3 & 94.1/97.8\&96.9/97.8 & 97.3/94.6\&-/-               & 98.3/96.9\&-/-               \\
                                                           & Switch                               & 91.9/86.2\&95.1/89.9                 & 94.9/91.3\&95.4/93.1                     & 88.8/89.7\&96.0/95.1 & 93.9/94.3\&95.5/95.6 & 96.5/95.2\&97.3/96.3 & 95.4/94.4\&-/-               & 96.9/94.8\&-/-               \\
                                                           & Tape                                 & 96.3/97.1\&98.5/97.9                 & 97.3/98.3\&98.7/98.6                     & 93.1/98.5\&98.3/99.7 & 96.8/98.7\&98.0/98.8 & 95.7/98.9\&98.6/99.3 & 98.1/97.8\&-/-               & 98.6/98.9\&-/-               \\
                                                           & Terminal Block                       & 91.4/92.8\&95.2/94.3                 & 96.0/97.3\&96.1/96.4                     & 85.7/99.3\&95.3/99.4 & 96.7/99.1\&97.8/99.0 & 94.9/99.0\&97.4/98.9 & 97.5/97.3\&-/-               & 97.9/99.0\&-/-               \\
                                                           & Toothbrush                           & 85.1/86.6\&84.8/85.9                 & 86.9/90.3\&86.9/88.3                     & 78.5/70.9\&79.1/69.4 & 78.1/89.5\&76.6/87.8 & 81.5/85.8\&81.8/84.5 & 83.0/85.1\&-/-               & 79.9/80.9\&-/-               \\
                                                           & Toy                                  & 75.9/82.1\&79.8/82.1                 & 83.5/87.0\&83.9/86.6                     & 79.1/83.6\&80.6/85.8 & 81.0/91.6\&85.9/92.0 & 81.0/90.0\&84.9/91.2 & 88.1/86.3\&-/-               & 80.1/81.8\&-/-               \\
                                                           & Toy Brick                            & 76.2/77.3\&76.7/77.6                 & 78.2/80.3\&79.1/81.1                     & 65.0/68.7\&73.1/79.2 & 68.5/78.3\&71.4/81.3 & 64.9/73.3\&70.3/77.6 & 78.6/71.9\&-/-               & 90.7/92.0\&-/-               \\
                                                           & Transistor1                          & 89.8/85.5\&94.4/92.4                 & 90.7/91.7\&95.9/96.2                     & 77.0/70.5\&89.1/86.9 & 91.6/97.3\&95.3/97.8 & 92.2/96.7\&95.1/97.1 & 93.7/94.9\&-/-               & 95.7/97.1\&-/-               \\
                                                           & USB                                  & 88.1/88.9\&92.3/92.7                 & 92.5/96.0\&93.2/95.7                     & 79.6/93.9\&94.9/98.4 & 93.1/98.3\&94.3/98.1 & 91.6/97.8\&95.3/98.4 & 95.0/97.2\&-/-               & 92.1/95.1\&-/-               \\
                                                           & USB Adaptor                          & 78.0/81.9\&82.2/82.3                 & 82.7/83.4\&85.7/84.3                     & 71.1/88.8\&87.3/97.0 & 79.0/90.6\&83.2/91.1 & 78.6/95.9\&86.0/96.3 & 89.1/86.2\&-/-               & 96.3/97.9\&-/-               \\
                                                           & U Block                              & 92.3/94.2\&93.4/94.8                 & 93.2/94.4\&93.5/96.0                     & 89.0/98.4\&94.5/99.0 & 93.1/98.6\&93.5/98.4 & 91.7/98.1\&94.4/98.5 & 93.5/95.8\&-/-               & 93.0/97.6\&-/-               \\
                                                           & Vcpill                               & 89.6/87.9\&90.8/88.9                 & 89.8/88.7\&90.5/88.3                     & 81.2/80.0\&94.1/94.1 & 87.1/89.8\&91.1/90.8 & 90.0/94.3\&93.0/94.8 & 92.5/91.7\&-/-               & 92.7/94.8\&-/-               \\
                                                           & Wooden Beads                         & 83.2/75.9\&88.4/82.3                 & 86.2/83.6\&88.7/85.6                     & 79.2/87.9\&90.1/87.2 & 83.5/90.3\&85.8/91.1 & 84.6/92.8\&88.7/93.8 & 90.5/90.3\&-/-               & 90.7/93.8\&-/-               \\
                                                           & Woodstick                            & 74.6/61.1\&78.5/67.9                 & 79.5/68.9\&80.3/71.0                     & 88.7/92.0\&87.7/94.2 & 86.3/90.6\&88.0/91.5 & 85.4/91.8\&89.5/92.0 & 89.7/89.9\&-/-               & 91.6/94.6\&-/-               \\
                                                           & Zipper                               & 94.7/92.2\&97.8/94.8                 & 95.0/94.2\&97.1/94.9                     & 91.0/88.8\&96.0/94.3 & 98.6/98.1\&99.6/98.4 & 97.0/96.8\&98.7/97.5 & 97.8/96.5\&-/-               & 97.6/96.9\&-/-               \\
            \hline
                                                           & Average All                          & 86.3/86.4\&88.8/88.4                 & 88.5/89.8\&89.3/90.2                     & 82.1/89.8\&89.3/93.0 & 87.8/94.3\&89.6/94.4 & 87.4/94.8\&90.5/95.3 & 90.9/91.1\&-/-               & 91.8/94.8\&-/-               \\
            \hline
        \end{tabular}
    }
\end{table*}

\end{onecolumn}


\end{document}